\documentclass[Afour,sagev,times]{sagej}

\usepackage{moreverb,url}

\usepackage[colorlinks,bookmarksopen,bookmarksnumbered,citecolor=red,urlcolor=red]{hyperref}
\usepackage{cleveref}
\usepackage{amssymb}

\newcommand\BibTeX{{\rmfamily B\kern-.05em \textsc{i\kern-.025em b}\kern-.08em
T\kern-.1667em\lower.7ex\hbox{E}\kern-.125emX}}

\newcommand{\add}[1]{\textcolor{black}{#1}}

\def\volumeyear{2024}

\begin{document}

\title{An Empirical Study of Counterfactual Visualization to Support Visual Causal Inference}


\author{Arran Zeyu Wang\affilnum{1}, David Borland\affilnum{2}, and David Gotz\affilnum{1}}

\affiliation{\affilnum{1} The University of North Carolina at Chapel Hill, Chapel Hill, NC, USA\\
\affilnum{2} RENCI, The University of North Carolina at Chapel Hill, Chapel Hill, NC, USA}

\corrauth{David Gotz, School of Information and Library Science, The University of North Carolina at Chapel Hill, Manning Hall, 216 Lenoir Drive, Chapel Hill, NC 27599, USA.}

\email{gotz@unc.edu}

\begin{abstract}
Counterfactuals---expressing what might have been true under different circumstances---have been widely applied in statistics and machine learning to help understand causal relationships. More recently, counterfactuals have begun to emerge as a technique being applied within visualization research. However, it remains unclear to what extent counterfactuals can aid with visual data communication. In this paper, we primarily focus on assessing the quality of users' understanding of data when provided with counterfactual visualizations. We propose a \add{preliminary model of} causality comprehension by connecting theories from causal inference and visual data communication. Leveraging this model, we conducted an empirical study to explore how counterfactuals can improve users' understanding of data in static visualizations. Our results indicate that visualizing counterfactuals had a positive impact on participants' interpretations of causal relations within datasets. These results motivate \add{a discussion of how to more effectively incorporate counterfactuals} into data visualizations.
\end{abstract}

\keywords{Counterfactual visualization, empirical study, causal inference, perception and cognition}

\maketitle

\section{Introduction}
\label{sec-intro}
Visualization has become an indispensable tool to help gain insights into increasingly large and complex multi-dimensional datasets.
However, effectively communicating meaningful causal relationships in such datasets remains challenging~\cite{gehlenborg2010visualization, kong2018frames, walny2019data}.
To help address this issue, researchers have explored the use of causal analysis theory to inform visualization design~\cite{wang2015visual, wang2017visual}.
One approach that has gained popularity is the use of \emph{counterfactual reasoning}~\cite{pearl2009causal, pearl2009causality}, a fundamental tool in statistical causal inference that uses hypothetical scenarios to investigate causal relationships. For example, an investigation of the effect of hunger on student test scores may involve considering the counterfactual where students ate lunch before sitting for an exam.
Counterfactuals have been applied in a number of visual analytics systems for machine learning explanation~\cite{cheng2020dece, gomez2020vice} and exploratory visual analysis~\cite{kaul2021improving}.

While some effective use cases have been reported in previous work, it remains unclear how and to what extent counterfactuals can help users gain a deeper understanding of visualized data.
Prior empirical studies in visual causal inference and counterfactual visualization focused on assessing spurious causal correlations~\cite{xiong2019illusion}, modeling treatment effects and confounding factors~\cite{kale2021causal}, and exploring users' confidence in feature-to-outcome relations~\cite{kaul2021improving}.
However, this existing body of work has mainly evaluated self-reported confidences and preferences in specific contexts. As a result, there remains a lack of general understanding of how applying counterfactuals to data visualization can benefit users' causal inferences.

In this paper, we explore whether and how counterfactuals in general-purpose visualizations can help users gain a deeper understanding of causal relationships within their data.
Drawing inspiration from research on the cognitive process behind visual data communication~\cite{adar2020communicative, ajani2021declutter, lee2022affective}, we first propose a novel visual causality comprehension model---capturing how we expect people to read and comprehend causal information from visualizations---that includes four progressive levels: recognizing, understanding, analyzing, and recalling.

Based on this \add{preliminary} model we conducted an empirical study employing four tasks derived from the corresponding comprehension levels.
\add{Motivated by prior work that examined how people can draw causal inferences from simple visualizations~\cite{xiong2019illusion, kale2021causal, kaul2021improving, bergstrom2018scatter},} three common visualization types---line charts, bar charts, and scatterplots---were used as stimuli to present various datasets.
For each dataset, a corresponding set of counterfactual-based visualizations was constructed based on the methodology used in prior work~\cite{kaul2021improving}.  
In each phase of our study, participants were randomly shown different combinations of chart sets displaying different levels of counterfactual information. They were then
asked to answer questions related to three design objectives: recognizing correlations, making predictions, and identifying causal relationships. Additionally, participants were asked to report how much they could recall about the datasets ten minutes after completing the study.

The study found that using counterfactuals in visualization design significantly improved participants' ability to understand and draw inferences from datasets, while also improving recall. 
Moreover, participants reported that counterfactuals helped them reason about hypothetical scenarios and identify causal relationships that were not immediately apparent.  The study results also suggest that counterfactual designs do not impair users' ability to read charts.
However, counterfactuals did require longer response times for answering questions.
Based on these findings, we propose a set of design heuristics to guide the integration of counterfactuals into data visualizations.
These guidelines can assist researchers and designers in creating effective counterfactual visualizations to aid in enhancing users' comprehension of complex data.

In summary, the contributions of this paper include:
\begin{itemize}
    \item {\bf A \add{preliminary model} of visual causality comprehension} that characterizes the human cognition processes used to understand visualizations in the context of causal inference theory.
    \item {\bf Results from an empirical study} evaluating the impact of counterfactual visualizations on the interpretation of datasets along a progression of communication levels.
    \item {\bf A set of design heuristics} to help guide future work with counterfactual visualizations in light of existing visual design guidelines.
\end{itemize}

\section{Related Work}
\label{sec-related}
This section introduces key definitions and provides an overview of relevant previous work, including prior research on visual data communication, the use of counterfactuals in visual analytics, and human visual perception of causality.

\subsection{Definitions for Counterfactual Visualization}

As briefly described in the introduction of this paper, \emph{counterfactual reasoning}~\cite{pearl2009causal, pearl2009causality} is a fundamental concept in statistical causal inference. This methodology involves analyzing what might happen under \emph{alternative scenarios} in which only a specific condition is different with the aim of understanding the causal effect of that specific condition. Many visual analytics workflows involve the creation of data subsets for focused analysis, as exemplified by the \emph{Zoom and Filter} step of Shneiderman's Mantra \cite{shneiderman1996eyes}. The counterfactual approach integrates well with such analyses, and can be operationalized through the creation of four different subsets: the \textbf{included (IN) subset}, the \textbf{excluded (EX) subset}, the \textbf{counterfactual (CF) subset} and the \textbf{remainder (REM) subset}~\cite{kaul2021improving}. These are defined as follows:

\begin{itemize}
\item \textbf{IN}: The data subset of interest, specified via certain inclusion criteria. 

\item \textbf{EX}: The rest of the dataset that has been excluded based on the inclusion criteria. This contains all data not included in IN.

\item \textbf{CF}:
Selected to include data points from EX that are similar to those in IN across all dimensions other than the inclusion criteria for IN. The CF subset therefore aims to enable counterfactual reasoning with respect to the data points in IN, as the user can investigate \emph{alternative scenarios} based on subsets that are similar to IN except for the fact that the inclusion criteria is not satisfied.

\item \textbf{REM}: The remaining data points from EX that are not included in CF. In other words, the excluded data points that are also dissimilar from IN.
\end{itemize}

\begin{figure}[htbp]
\centering
\includegraphics[width=1\columnwidth]{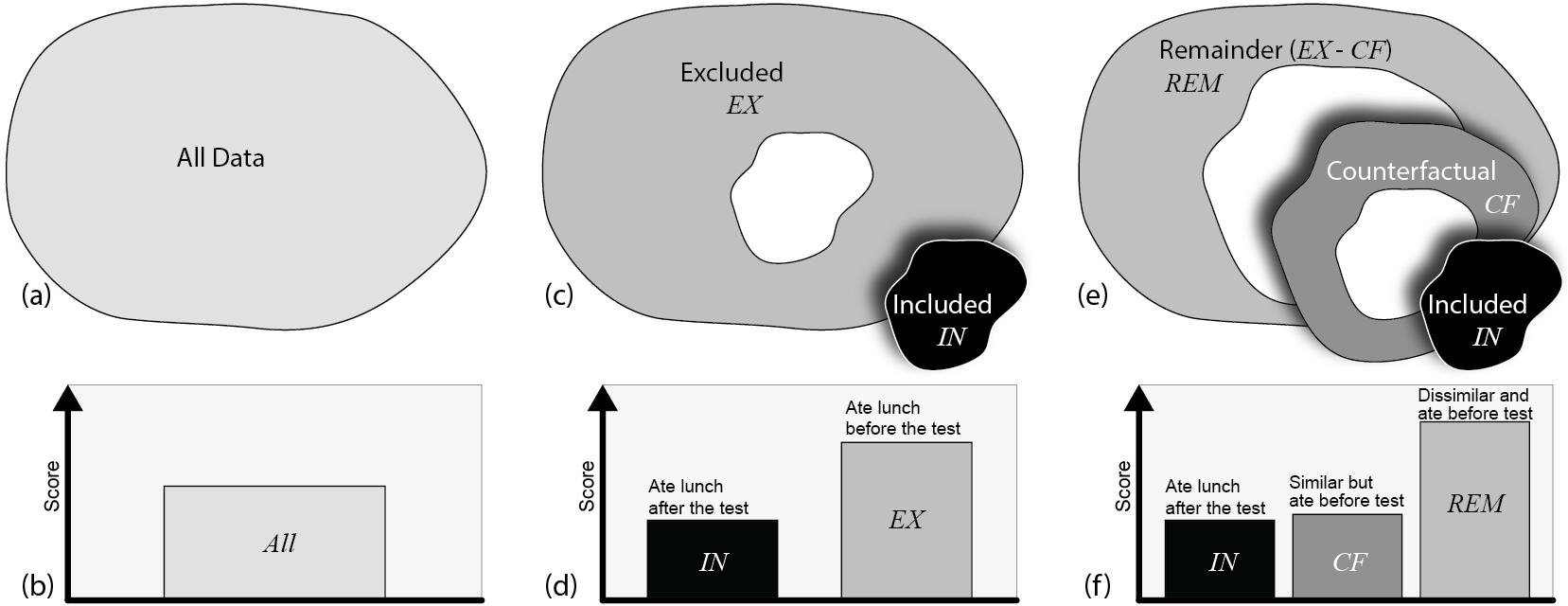}
\caption{The four types of data subsets used in our study, illustrated with the student test scores example from the introduction.
(a-b) are all data points in the dataset and corresponding traditional bar chart visualization of the average test score for all students.
When the students who ate launch after the test are selected as IN, (c-d) shows the subset relations and visualizations between the IN and EX subsets, and (e-f) shows the relations and visualizations between the IN, CF, and REM subsets.
}
\vspace{-1em}
\label{fig:subsets}
\end{figure}

These subsets are illustrated in \autoref{fig:subsets} (a), (c), and (e), and the method for selecting them for the purposes of the user study presented in this paper is described in \nameref{sec-stimuli}.

Counterfactual visualization aims to provide comparisons of useful combinations of data subsets that can provide additional information to improve reasoning about causal relationships in the data.
Following Kaul et al.~\cite{kaul2021improving}, we refer to \emph{a counterfactual visualization} as a combination of three charts showing the IN, CF, and REM subsets (\autoref{fig:subsets} (f)). A traditional visualization of a chart showing just the IN subset is treated as a control group in our study (e.g., a scatter plot showing positions of only data points in IN). 

However, since counterfactual visualizations include a comparison across multiple charts (IN + CF + REM), a second control group with two charts showing the IN and EX subsets 
(e.g., \autoref{fig:subsets} (d)) is also included in the study design reported in this paper.
The charts for this second control group display all data points as is the case with counterfactual visualization designs, but they only show the EX subset rather than the similarity-driven subsets CF and REM. 

The purpose of visualizing CF is to show the user data points that are similar to IN across all dimensions in the data other than the inclusion criteria for IN, thus helping them confirm or deny any causal relationships they may assume from looking at IN alone, based on the inclusion criteria for IN. E.g., in the student test scores example from the introduction, if IN contains students who ate lunch after taking a test, and a visualization shows that they have low test scores, users might assume a causal relationship between hunger and test scores (see \autoref{fig:subsets} (d)). In this case, CF would contain students who ate lunch before taking the test, but who are similar to IN in all other respects. If the students in CF also have low test scores, that would weaken the evidence for a causal link between hunger and test scores (see \autoref{fig:subsets} (f)). On the other hand, if the students in CF have high test scores, that strengthens the evidence of a causal relationship. The EX and REM subsets contain individuals dissimilar to IN across all dimensions of the data, providing further context to help with causal inference.
\autoref{fig:subsets} (b), (d), and (f) show visualizations for the student test scores example.

\subsection{Human Cognitive Processes and Visual Data Communication}
\label{sec-vdc}

One stated advantage of visualization is that it ``forces us to notice what we never expected to see''~\cite{tukey1962future, tukey1977exploratory}  within datasets quickly and easily~\cite{van2005value, munzner2014visualization}.
Designers aim to present information to users in the most effective way.
However, achieving these goals is not always easy, as the efficiency of visualizations can be influenced by various factors such as visual encodings, data type, and even designers' own biases~\cite{kong2018frames, lee2022affective}.

Existing research has examined various factors impacting the comprehension and communication of data in visualizations.
Task-based analyses are a common approach, in which researchers explore and summarize how to design visualizations to aid understanding for specific tasks, such as low-level graphical perception~\cite{cleveland1984graphical, heer2008graphical} and color design~\cite{szafir2018modeling, tseng2023measuring}.
In addition to task-based studies, researchers have also investigated how to assess and model users' understanding and ability to communicate visualizations through evaluating study strategy design~\cite{sedlmair2012design, sedlmair2015data}, and visual quality measures~\cite{wilkinson2005graph, wang2019improving}.
Interdisciplinary insights have also been proposed, such as Bae et al.'s assessment of how curiosity and play in physicalizations improve data visualization literacy in children's education~\cite{bae2022cultivating}.

Moreover, Adar and Lee~\cite{adar2020communicative, lee2022affective} built on previous studies to develop an affective learning objective framework that aligns with Bloom's affective taxonomy~\cite{bloom2020taxonomy}.
They conceptualized the visual data communication problem between designers and users as a learning problem in a teacher-student relationship.
By doing so, they summarized insights in human cognition objectives that enable designers to describe their visualizations' goals and compare their communication ability with users to those goals in a formalized way.

In this paper, we gain insights from prior insights on human cognition objectives in visual data communication to counterfactual visualization.
We aim to explore how people comprehend counterfactuals in visualizations, taking into account the specific demands of counterfactual reasoning.

\subsection{Counterfactuals in Visual Analytics}
\label{sec-cf-va}

Counterfactual reasoning is a fundamental concept in statistical causal inference~\cite{pearl2009causal, pearl2009causality}.
It involves constructing hypothetical scenarios that deviate from reality and making inferences about what would happen under those counterfactual conditions.
For example, we might ask, ``What would the sales figures have been if we had lowered our prices by 10\% last year?'' to assess the causal relations between price and sales figures. Counterfactual reasoning has been widely recognized for its importance
and has recently gained traction in the deep learning community, where it has been applied for tasks like model testing~\cite{wu2021polyjuice} and narrative reasoning~\cite{qin2019counterfactual}.
While most of these studies are non-visual and result in natural language output, we focus here on the application of counterfactual reasoning in the context of visual analytics~\cite{borland2024using}.

The vast majority of existing work on counterfactual visualizations has focused on improving explanations for and the interpretability of machine learning models.
For instance, the \emph{What-If Tool}~\cite{wexler2019if} provides a basic visualization of the nearest counterfactual point to the target data point, 
\emph{ViCE}~\cite{gomez2020vice} uses counterfactuals to illustrate minimal edits required to modify the output of the visualized model,
\emph{DECE}~\cite{cheng2020dece} enables the visualization of counterfactual examples from different data subsets for decision-making, and \emph{INTERACT}~\cite{ciorna2023interact} allows what-if analysis to improve model explainability and prototyping under industrial contexts.
Although these studies present effective use cases for their systems, they focus primarily on machine learning model explanations rather than providing insights for general-purpose visualizations.

The potential uses of counterfactuals are not limited to the problem of explaining machine learning models~\cite{borland2024using}.
In recent work, Kaul et al.~\cite{kaul2021improving} presented \emph{CoFact}, the first counterfactual-based interactive visualization system designed for general datasets.
\emph{CoFact} enables users to explore high-dimensional datasets \add{via pairwise visualization of features, prioritized based on association with an outcome variable. Counterfactual visualizations of selected data subsets provide additional information about feature-to-outcome relationships.}
Although this work provided initial findings on the usefulness of counterfactual visualization in the context of an interactive visual analytics tool, there remains a gap in understanding the impact of counterfactuals for more general data visualization and communication.

Kaul et al.~\cite{kaul2021improving} further assessed how the \emph{CoFact} system helped users learn feature-to-outcome relationships from datasets.
Their results provided preliminary evidence that, with counterfactual visualization enabled, users showed greater confidence in strong outcome relationships and lower confidence in weak outcome relationships.
The post hoc analysis of interviews found that \emph{CoFact} could be useful for data exploration and decision-making.
However, their study mainly focused on the proposed visual analytic system and lacked insights for more general visualizations.
Furthermore, they did not assess the quality of causal inferences generated by users with the help of counterfactuals. The study in this paper aims to address both of these issues.

\subsection{Human Visual Perception of Causality}
\label{sec-causal-study}

Properly designed visualizations can help users avoid making spurious assumptions about causal relationships, leading to improved decision-making~\cite{oghbaie2016understanding}.
Thus, understanding how human perceptions of causal inference are processed and impacted by visualizations is important for visualization research.

Xiong et al.~\cite{xiong2019illusion} explored how various graphs can create an illusion of causality, i.e., how they can lead to incorrect interpretations of data.
They introduced some preliminary insights for visualization design, such as causal inference results for text descriptions and bar graphs being better than those for scatterplots.
Another crucial finding from their work is that the data aggregation level of visualizations might be positively associated with users' self-reported confidence in causal inferences.
However, their study focused on assessing users' self-reported ratings of pre-designed causality statements, thus lacking insights of users' actual causal inference results.

Further, Kale et al.~\cite{kale2021causal} introduced an empirical study for evaluating causal inferences via the causal support model from mathematical psychology~\cite{griffiths2005structure}.
Their results indicate that user capability for causal inferences is insensitive to sample size.
They also reported that using different visual encodings would not be significantly better than tables for causal inference.
However, their second finding is inconsistent with most existing visual causal analytics systems all of which indicate that visualization would benefit causal inference~\cite{deng2021compass, xie2020visual, cheng2020dece, gomez2020vice, kaul2021improving, guo2023causalvis}.
This difference may be related to different design strategies between the empirical study and interactive visual analytics systems, but still needs to be further studied.

\add{Network visualizations such as Bayesian Belief Networks~\cite{lam1994learning} are often employed for causal inference tasks. 
However, a number of prior studies~\cite{kaul2021improving, kale2021causal, xiong2019illusion, bergstrom2018scatter} have shown that users also draw causal inferences from common visualizations including scatterplots, line charts, and bar charts, even if they were not intended to show causal relationships.
Moreover, starting from simplified and easy-to-understand tasks is a common and important approach to exploring complex concepts in empirical studies~\cite{plaisant2004challenge, lam2011empirical}. For these reasons, our study primarily focused on these three common visualization types.}

In this study, we aim to systematically evaluate and model the impact and effectiveness of counterfactuals in helping users understand data at different data communication levels~\cite{adar2020communicative} for general-purpose charts, building upon the previous definition of counterfactual visualization~\cite{kaul2021improving}.
Compared to existing studies, we primarily focus on evaluating the quality of users' causal inference results instead of just assessing self-reported confidence or preference levels, and discuss the design space of how to use and understand counterfactuals in visualization.
By doing so, we intend to provide empirical evidence supporting the use of counterfactuals in visual analytics.

\section{Modeling Causality Comprehension}
\label{sec-model}
Existing empirical studies on causality in visualization~\cite{xiong2019illusion, kale2021causal} have not explored the perceptual data communication process underlying visual causal inference.
To advance our understanding of how counterfactuals can enhance users' comprehension of data, we propose a \add{preliminary} causality comprehension model for visualization scenarios.
The proposed model aims to decompose this process and shed light on the potential benefits of using counterfactuals in data communication.

According to statistical causal inference theory~\cite{pearl2009causal, pearl2009causality}, causalities can be classified into three levels: association, intervention, and counterfactual.
In this paper, we propose a model of users' progressive understanding of causalities in visualization by connecting these three causal inference levels with four important cognitive objectives that occur when users communicate with visualizations~\cite{heer2008graphical, adar2020communicative, lee2022affective, borkin2015beyond} --- \emph{Recognize}, \emph{Understand}, \emph{Analyze}, and \emph{Recall} --- resulting in four comprehension levels:

\begin{figure}[htbp]
\centering
\includegraphics[width=0.9\columnwidth]{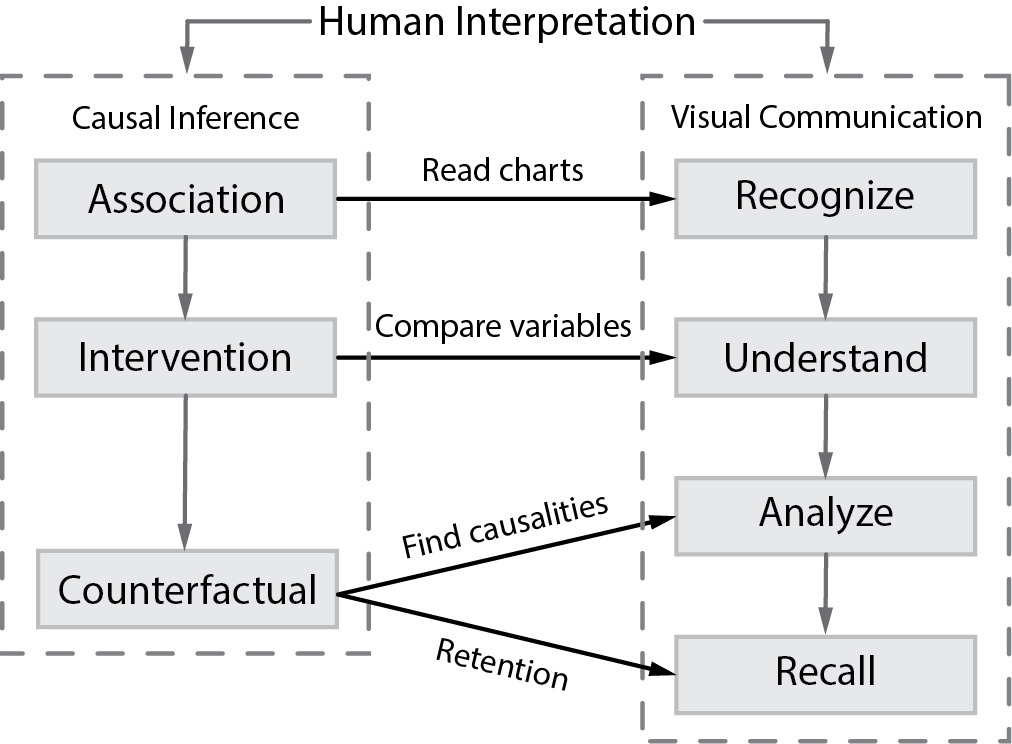}
\caption{Framework of the proposed causality comprehension model. The left dashed box shows the causal inference theory~\cite{pearl2009causal, pearl2009causality}, connected to cognitive objectives in visual data communication on the right.}
\label{fig:model}
\end{figure}

\begin{itemize}
    \item {\bf Association $\rightarrow$ Recognize:}
    At the preliminary level of causal inference, \emph{association} involves identifying statistical correlations between variables~\cite{pearl2009causality, pearl2009causal}, e.g., ``what does this survey tell us about the election results?''
    Such correlations can be directly expressed in a simple chart, e.g., showing the monotonic relation of two axes in a scatterplot.
    Users can typically identify these correlations by directly reading a chart.
    This ability is closely related to the \emph{recognize} process in human cognition.
    
    \item {\bf Intervention $\rightarrow$ Understand:}
    The second level, \emph{intervention}, involves manipulating one variable to observe the effect on another variable in a dataset, e.g., ``will my headache be cured if I take aspirin?''
    This level requires users to interpret the meaning of variables, summarize and compare their trends, and make relevant predictions.
    These aspects are expressed as the cognitive ability to \emph{understand}.

    \item {\bf Counterfactual $\rightarrow$ Analyze:}
    The highest causal inference level, counterfactual, involves predicting what would have happened if a different intervention had been made in the past, e.g., ``what if I hadn't gone to college in the past?''
    Counterfactual thinking involves thinking about the impact of other related variables in this dataset.
    It is more complex than intervention and requires distinguishing the interaction effects of different variables, integrating cross-variable insights, and deconstructing their impact across the whole dataset. 
    We therefore connect it to an advanced cognitive ability in visual data communication---\emph{analyze}.

    \item {\bf Counterfactual $\rightarrow$ Recall:}
    In addition to the above connections, we include the cognitive ability of \emph{recall}, which describes the memorability of visual communication and is a component complementary to \emph{recognize}~\cite{adar2020communicative}.  
    We therefore also connect \emph{counterfactual}  with  \emph{recall}.
    We placed it as the last step of causality comprehension because recall appears in the final stage of visual understanding in general~\cite{borkin2015beyond, kong2019trust}.
    
\end{itemize}

In summary, we define a \add{preliminary model of the} visual causality comprehension process of human perception as a progression from recognizing, to understanding, to analyzing, and finally recalling, connecting the theory of causal inference and users' cognitive processes of visual data communication.
\autoref{fig:model} illustrates an overall picture of our model.

\section{Methodology}
\label{sec-method}

To assess how counterfactual visualizations impact people's ability to reason about and comprehend causality in real-world datasets, we ran a user study that was approved by the [Redacted] Institutional Review Board.
This study enabled us to characterize the effect of counterfactuals in different types of static visualizations, including to what extent they help users infer causal relationships.
The datasets and example infrastructure applied in our study are available in the supplemental material.

\subsection{Participants}
We recruited 32 participants (\add{19 male and 13 female}, based on a power analysis~\cite{faul2007g} of pilot studies) via recruitment flyers, department mailing lists, and contacts within professional networks.
All participants were at least 18 years old, reported normal or corrected to normal vision, and were either pursuing or had earned a university degree. 
Participants were from a broad spectrum of academic and professional domains. 
Our experiment took 45 minutes on average, and each of the participants was compensated \$10 for their time.

\subsection{Hypotheses}

Based on the proposed causality comprehension model, we hypothesized that:

\begin{itemize}
\item[\textbf{H1:}] \textbf{Counterfactuals would not hurt people's ability to recognize features of data.}

As the most basic level of communication, the recognizing process always appears in low-level visual tasks~\cite{amar2005low}.
Visual complexities and design choices within a chart could impact its performance~\cite{regier2009language, quadri2021survey}. However, previous work suggests that counterfactual visualizations can be integrated into a visual analytics system without decreasing system usability \cite{kaul2021improving}. In addition, we chose a \emph{juxtaposition} visual comparison model that has been employed by many existing studies \cite{gleicher2011visual, zhang2015glyph, gleicher2017considerations} to maintain a low visual complexity for visualizations of each data subset (\autoref{fig:instance} (b) and (c)).

\item[\textbf{H2:}] \textbf{Counterfactuals would help people's understanding of datasets.}

In existing visual analytic systems~\cite{cheng2020dece, gomez2020vice, kaul2021improving}, counterfactuals are shown to be effective in understanding complex algorithms and judging feature-to-outcome relationships.
We expect that counterfactuals will also be beneficial to help understand data using general-purpose charts.

\item[\textbf{H3:}] \textbf{Counterfactuals would help people better find and analyze causal relationships in datasets.}

An obvious advantage of counterfactuals is to make apparent underlying causalities in data~\cite{pearl2009causality, kale2021causal}.
We expect their impact to be similar to that of insight explorations which also aim at finding hidden relations in data.
Insight exploration methods have been demonstrated to be useful in deeper data analysis~\cite{tang2017extracting, ding2019quickinsights, ma2021metainsight}.
Such findings indicate that we may see similar advances with counterfactuals, i.e., improved analyses and causal inferences.

\item[\textbf{H4:}] \textbf{Counterfactuals would improve the performance of recalling data.}

Perceptual studies demonstrate that additional information and encodings can help people better communicate and recall data~\cite{borkin2015beyond}.
Counterfactual visualizations provide extra information to users, and we anticipate that such counterfactual information will help people remember and recall findings from data more easily.

\end{itemize}

\subsection{Stimuli}
\label{sec-stimuli}

Our stimuli were created from commonly used real-world multi-dimensional datasets found in prior studies, such as the \emph{UCI Credit Card dataset}\footnote{https://archive.ics.uci.edu/ml/datasets/default+of+credit+card+clients} \cite{yeh2009comparisons} and the \emph{Census Income dataset}\footnote{https://archive.ics.uci.edu/ml/datasets/adult} \cite{kohavi1996scaling}. We followed the process of the \emph{CoFact} system~\cite{kaul2021improving} for generating the data subsets for visualization, as it is currently the only counterfactual-based system for exploratory visual analysis.

We generated the data subsets as follows:
\begin{itemize}
    \item Picking an included (IN) data subset based on the variables of top-recommended insights computed by a dataset insight exploration algorithm~\cite{tang2017extracting}, e.g., all individuals with one child from a family-income dataset. This selection also results in an excluded (EX) data subset (individuals that do not have one child).

    \item Creating the counterfactual (CF) data subset from EX following previous work~\cite{kaul2021improving} by (i) computing the Euclidean distance from each point in EX to each point in IN, and (ii) selecting the $n$ points from EX that have the shortest total distance, where $n = |IN|$.

    \item Denoting the remaining data (neither in IN nor CF) as the remainder (REM) data subset, resulting in four subsets (IN, EX, CF, and REM) for each dataset (See \autoref{fig:subsets}).
\end{itemize}

\begin{figure*}[htbp]
\centering
\includegraphics[width=1.8\columnwidth]{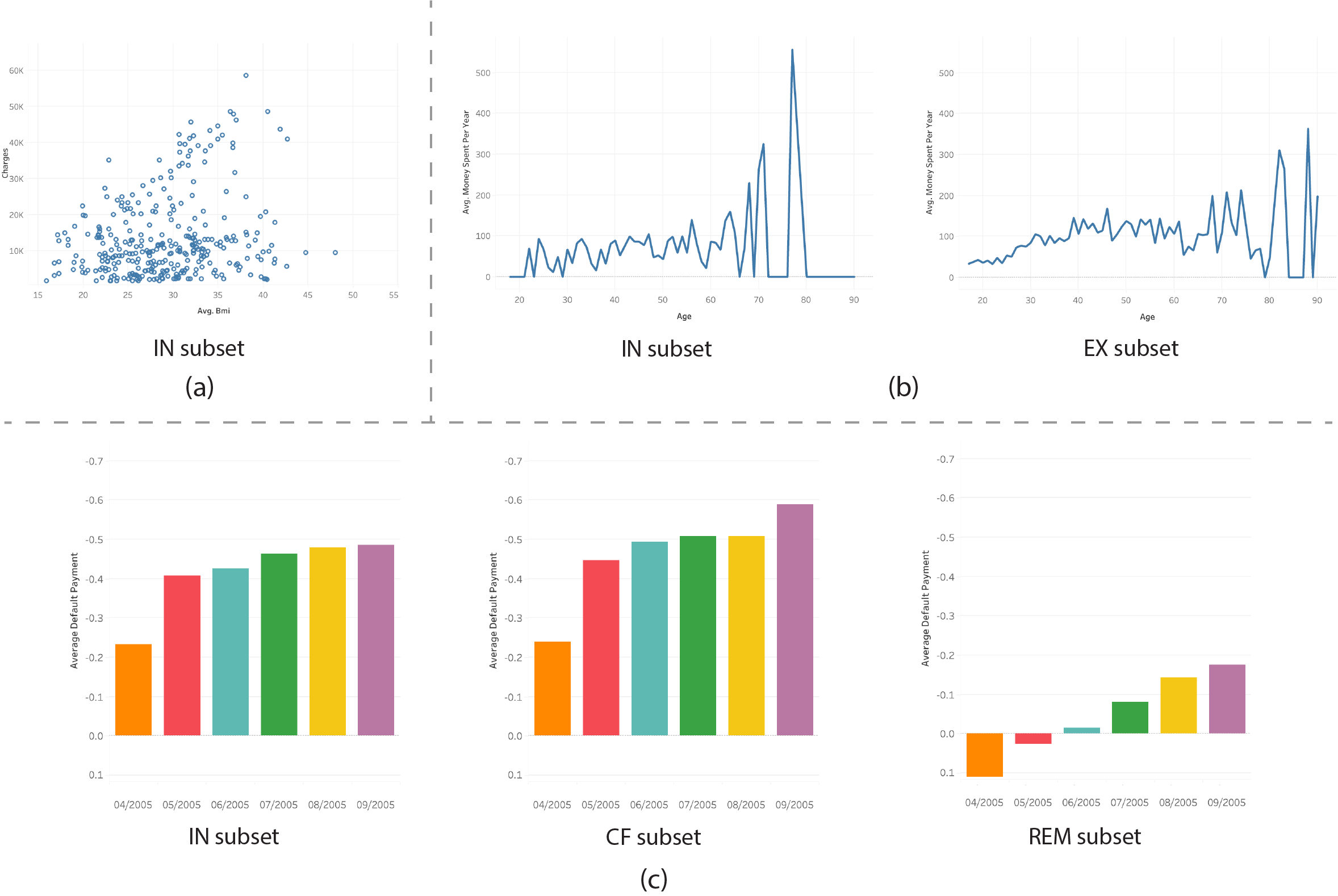}
\caption{Three examples of data subset visualizations seen by participants in the user study:
(a) IN subset scatterplot visualization of the \emph{Health Insurance} dataset, showing only data in the IN subset.
(b) EX subset line charts visualization of the \emph{Census Income} dataset, showing data in both the IN and EX subsets.
(c) CF subset bar charts visualization of the \emph{UCI Credit Card} dataset, showing data in the IN, CF, and REM subsets.
}
\label{fig:instance}
\end{figure*}

The stimuli were separated into three groups containing different subsets to be visualized: 
\begin{itemize}
    \item  IN group---IN subset only (e.g., \autoref{fig:instance} (a)),
    \item EX group---IN + EX subsets (e.g., \autoref{fig:instance} (b)), and 
    \item CF group---IN + CF + REM subsets (e.g., \autoref{fig:instance} (c)).
\end{itemize}

We chose three common visualization types to display the data subsets: line charts (for time-series data), bar charts (for categorical data), and scatterplots (for continuous data).
In total, we generated 27 groups of visualizations and employed them in a within-subject study.
\autoref{fig:instance} shows examples of the three groups of data subset visualizations with different chart types.

\subsection{Tasks}
\label{sec-tasks}

We derived four tasks based on our causality comprehension model:

\begin{itemize}
\item[\textbf{T1:}] Describe anything of interest noticed by looking at the current visualization.

\item[\textbf{T2:}] Predict changes in a variable if a different variable were to be manipulated.

\item[\textbf{T3:}] Make broader predictions about what will happen if a particular variable from the current chart were to be changed or replaced with another one.

\item[\textbf{T4:}] Report recall of tested visualizations.

\end{itemize}

\begin{table}[htbp]
\centering
  \caption{Examples of questions per task.}
  \label{tab:questions}
  \begin{tabular}{c|p{7cm}}
    \hline
    \textbf{Task} & \textbf{Example Question}\\
    \hline
    T1 & Look at the charts and describe anything you can recognize from the visualization. \\
    T2 & What will the remaining loan value change (increase, decrease, or remain similar) if people's credit limits become higher? \\
    T3 & What will the data change in the above-shown chart if people's marital status changes to divorce? For example, think about average values, trends, and distributions. \\
    T4 & Describe visualizations that you can still remember. \\
  \hline
\end{tabular}
\end{table}

Participants were required to answer three questions for each task from T1 to T3, and were shown each subset combination group (as introduced in \nameref{sec-stimuli}) once for each task.
For questions asked in T2, we hid the last 5\% of the dataset to serve as the ground truth, following the approximate percentage for the validation set in the Microsoft COCO dataset~\cite{lin2014microsoft}, to use as validation for users' answers.
\autoref{tab:questions} shows examples of specific questions asked for the different tasks.

\subsection{Procedure}
Our experiment consisted of five phases: (1) informed consent, (2) term introduction and task description, (3) formal study for T1-T3, (4) post-study feedback and preference questionnaire, and (5) answering the recall question for T4.

Participants were shown and agreed to the informed consent with our IRB protocol at the beginning of the study.
We then explained any unfamiliar terms and provided examples appropriate for a general audience---we explained the definition of different subsets, provided examples of counterfactuals with narrative explanation (similar to examples in \nameref{sec-cf-va}), and encouraged users to imagine hypothetical assumptions during the study---before introducing the required tasks in the study. 

After completing the study introduction, each participant was required to view visualizations and answer questions for tasks T1-T3 in a random order, while avoiding back-to-back questions for the same task, to account for learning effects.
Participants provided their answers in a textbox.
In addition, they provided their confidence for each answer via a 5-point Likert scale.
During the study, each participant completed 3 questions for each task---1$\times$IN group, 1$\times$EX group, and 1$\times$CF group---and was shown a particular data set + visualization group combination once.
See \nameref{sec-stimuli} for definitions of the three subset groups.

After completing T1-T3, each participant completed a questionnaire reporting their experience including usefulness, confidence, preference, and any additional feedback.
We extended the feedback session by 10 minutes by chatting with participants and finally asked them to report their recall of any information remembered from the viewed visualizations.
The recall list was ranked from the most memorable to the least memorable by each user.

\subsection{Result Encoding}

We collected participants' thinking and completion time, verbal and textual responses, and reported confidence in each question.
For natural language input, we encoded them as quantities using axial coding~\cite{charmaz2006constructing} (see \autoref{tab:encoding} for details about the evaluation metrics).
We received 320 responses in total, with 288 for tasks T1 to T3, and 32 responses for task T4.
The responses were encoded into four different types following our proposed tasks and the model defined in \nameref{sec-model}.

\begin{table}[htbp]
\centering
  \caption{Evaluation metrics for each task.}
  \label{tab:encoding}
  \begin{tabular}{c|p{7cm}}
    \hline
    \textbf{Task} & \textbf{Evaluation Metric}\\
    \hline
    T1 & Number of findings, correctness rate \\
    T2 & Correctness rate \\
    T3 & Correctness rate, relative impact ratio (see \nameref{sec-res-level3} for its mathematical definition) \\
    T4 & Number of recalled datasets \\
  \hline
\end{tabular}
\end{table}

For task T1, we extracted the statistical descriptions and findings from reported responses, computed the number of findings in each response, and verified the correctness of the findings.
Note that since T1 was focused on the ability to recognize, users were requested to describe findings without inferring any correlations between variables.
For T2, we collected the prediction results and validated their correctness based on the remaining 5\% of the dataset, as described in \nameref{sec-tasks}.
For T3, we collected findings and predictions from users' responses and computed the average correctness.
Meanwhile, for both T2 and T3, we also collected the evidence or reason to support the answer to each question if it was reported.
Finally, we recorded the number of recalled datasets per visualized subset type from each response for T4.
\autoref{tab:encoding} describes the selected evaluation metrics for each task.

\section{Results}
\label{sec-results}

We present our analysis methodologies, statistical analysis, and significant results based on the independent factors considered in this paper (see \nameref{sec-analysis}) using both traditional inferential measures and 95\% bootstrapped confidence intervals ($\pm$ 95\% CI) for fair statistical communication~\cite{dragicevic2016fair}.

\subsection{Analysis}
\label{sec-analysis}

The overall goals of our analysis were to test the proposed hypotheses and to validate related findings from previous studies. To achieve these goals, 
we analyzed performance using the following quantitative evaluation metrics: completion time, reported confidence, number of reported findings or variables, correctness rate, and recall rate.
For each task, we assessed the resulting data using a 3 (visualized subset groups: IN, EX, and CF) factors ANOVA, where the chart types, inter-participant differences, and trial order are treated as random covariates, with Tukey's honestly significant difference test (Tukey's HSD) with $\alpha$ = 0.05 and Bonferroni correction as post hoc analysis. 
Further, we explored details of participants' qualitative responses to find additional potential insights into the effects of counterfactual visualizations.

\begin{figure}[htbp]
\centering
\includegraphics[width=0.8\columnwidth]{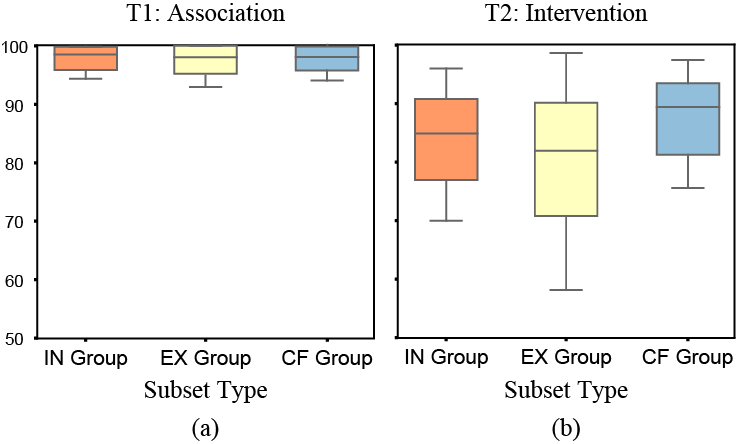}
\caption{The box plots show correctness rates for each visualized subset group for tasks T1 (a) and T2 (b).}
\label{fig:t12}
\end{figure}

\subsection{Recognizing Association}
\label{sec-res-level1}

Our results for task T1 support H1: we found that additional visualized subsets do not have an obvious negative impact on users' recognition ability.

During this task, users were asked to focus on findings from the data shown in the IN subset chart.
Our analysis shows that there is no significant impact between the visualized subsets and users' recognition results for the number of findings and correctness rate.
We found users always responded to similar findings for the same dataset.
For example, in the $CO_{2}$ emission dataset~\cite{owidco2andgreenhousegasemissions}, one user answered ``\emph{The CO2 emission goes higher and then reaches peak in 10-ish years and goes down.}'' after seeing the IN group and another user answered ``\emph{The emissions subsequent climb through the end point of 2010, and then slowly go back down.}''  after seeing the CF group.
Additionally, the overall correctness rate is near 100\% regardless of visualization type, as shown in \autoref{fig:t12} (a).

\subsection{Understanding Intervention}
\label{sec-res-level2}

Our results for T2 may also support \textit{H2}: counterfactual visualizations could improve users' interpretation of interventions behind data variables.

No overall significant difference was found between visualized subsets for users' correctness rate when predicting changes to a variable after manipulating another variable.
\autoref{fig:t12} (b) shows the correctness rate of users' predictions based on subset visualization type.
Although the overall statistical significance was not found at the 0.05 level, a higher average correctness rate for the CF group (IN + CF + REM) was observed visually, as was a larger variance for the EX group.

We therefore further explored the statistical significance at a more fine-grained level.
We found a significant difference ($p = 0.01, \eta^{2} = 0.07$) when only comparing the results from the IN and CF groups.
It could suggest that counterfactuals may have the potential to communicate variable relations more effectively for users by comparing against the IN subset, whereas the EX subset may introduce more random effects to such judgments. Further study is necessary to confirm or deny these relationships, however.

\subsection{Analyzing Causality}
\label{sec-res-level3}

Our results for T3 support H3: we found that counterfactual visualizations significantly improved users' analysis of causalities.

Our analysis reveals a significant impact between visualized subsets and users' overall correctness rate of causality results ($F(2, 24)=8.71, p = 0.001, \eta^{2} = 0.12$).
Other than the correctness rate, we further evaluated the results using a relative impact ratio; the relative impact ratio $r$ for user $i$ for a specific question $x$ is defined as:
\begin{equation}
    r_i(x)=\frac{Num_i(x)}{\mathop{\max}\limits_{j \in users}(Num_j(x))},
\end{equation}
where $Num_i(x)$ is the number of correct predictions in user $i$'s response and $\mathop{\max}\limits_{j \in users}(Num_j(x))$ refers to the maximum number of correct predictions made among every user that answered this question.

\begin{figure}[htbp]
\centering
\includegraphics[width=0.8\columnwidth]{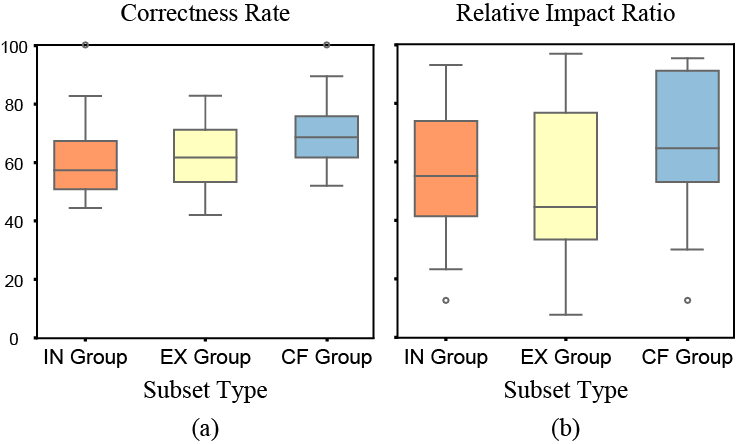}
\caption{The box plots show correctness rates (a) and relative impact ratios (b) of each visualized subset group for task T3.}
\label{fig:causality}
\end{figure}

\autoref{fig:causality} shows the results for correctness rate and relative impact ratio per visualized subset group.
Counterfactual visualizations achieved the highest average correctness (\autoref{fig:causality} (a)) and relative impact ratio (\autoref{fig:causality} (b)). The distribution of the CF group's results is also more compact compared to the EX group, implying that counterfactuals may be able to improve user's causal inference, whereas the EX group may in fact be a hindrance (similar to the results of T2 in \autoref{fig:t12}).
Although we do not specifically test these hypotheses in our study, our results could provide guidance on which combinations of data subsets to present to users.

\subsection{Recall}
\label{sec-res-level4}

Our results for task \emph{T4} support \textit{H4}: we found that counterfactual visualizations led to better recall rates for users.

We recognized all responses linked to a specific dataset as a successful recall; descriptions that could not be associated with a specific dataset were not recognized as a recall.
For example, one user said ``\emph{I remember poor countries with life expectancy vs money,}'' which was recognized as a recall because it could be directly linked to the life expectancy dataset~\cite{world2021ghe}, while another user said ``\emph{I remember the scatterplots which are easiest to use when visualizing data,}'' which we did not recognize as a recall because it could not be associated with a specific dataset.

\begin{figure}[htbp]
\centering
\includegraphics[width=0.6\columnwidth]{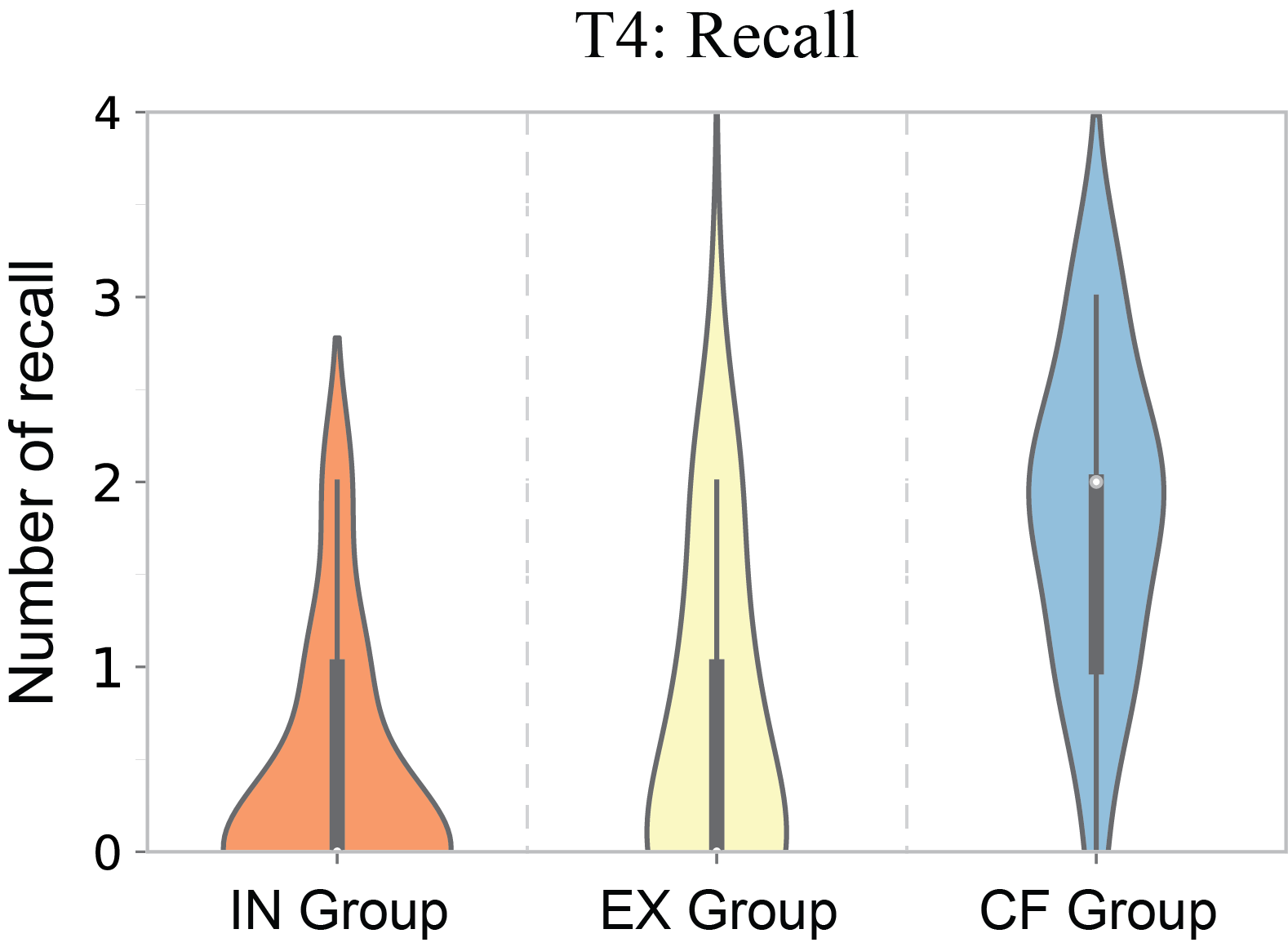}
\caption{The violin plots show the numbers of recalls per visualized subset group type.
}
\label{fig:recall}
\end{figure}

We found a significant impact between visualized subsets and users' recall of datasets ($F(2, 24)=1.12, p < .0001, \eta^{2} = 0.36$).
\autoref{fig:recall} shows the average recalled numbers of datasets per visualized subset group.

As shown in \autoref{fig:recall}, the CF (avg. 1.78 per user) group had a higher average number of recalls, followed by EX (avg. 0.69 per user) and IN (avg. 0.38 per user).
This finding could be due to counterfactuals causing users to perform a more careful causal analysis of data~\cite{kaul2021improving}. The effect could also be due to the additional information (for both EX and CF groups), which is in line with findings from Borkin et al. ~\cite{borkin2015beyond}.

\subsection{Exploratory Analysis}
\label{sec-exp}

To better analyze other potential impacts in our study, we conducted an exploratory analysis using Tukey’s HSD with Bonferroni correction of the other evaluation metrics and random covariates. Here we report significant results from this analysis.

\begingroup
\renewcommand{\arraystretch}{1.5}
\begin{table}[htbp]
    \centering
    \caption{The average response time (seconds) per visualized subset group type for T2 and T3.}
    \begin{tabular}{|c|c|c|c|}
    \hline
    Visualized Subset Group & IN & EX & CF \\
    \hline
    Response Time for T2 (sec) & 72 & 117 & 155 \\
    \hline
    Response Time for T3 (sec) & 131 & 170 & 232 \\
    \hline
    \end{tabular}
    \label{tab:time-p}
\end{table}
\endgroup

First, we report that counterfactuals led to an impact on the response time for users.
Our results reveal significant impacts on tasks T2 ($F(2, 24)=29.56, p < .0001, \eta^{2} = 0.51$) and T3 ($F(2, 24)=36.44, p < .0001, \eta^{2} = 0.47$) of visualized subset types on the response time.
\autoref{tab:time-p} provides the average response time of T2 and T3 for the three visualized subset groups.
This finding indicates that visualizing counterfactuals may lead to longer response and analysis times for users in reading charts.
We assume this is because the CF group introduced additional charts and information which require users to think deeply and more carefully compared to the original IN group~\cite{kaul2021improving}.
However, future work should be conducted more systematically to assess this hypothesis.

\begin{figure}[htbp]
\centering
\includegraphics[width=1\columnwidth]{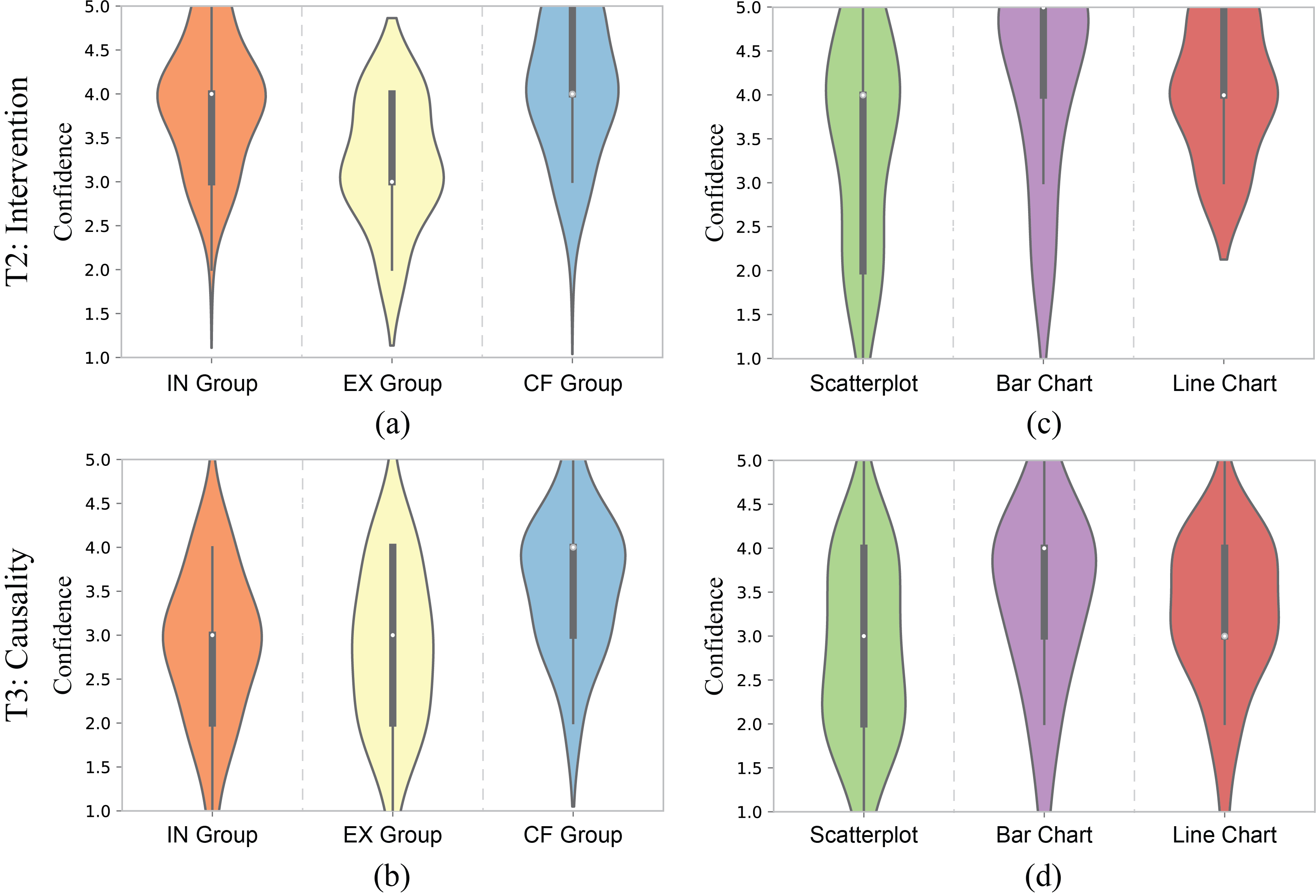}
\caption{The violin plots show the confidence (reported on a 5-point Likert scale) per visualized subset group for T2 (a) and T3 (b) and per chart type for T2 (c) and T3 (d).
}
\label{fig:confidence-vis}
\end{figure}

In addition, we found most of the users' reported confidences are at a moderate level, but they varied for both visualized subsets and chart types.
The results in \autoref{fig:confidence-vis} (a) and (b) show users' average confidence per visualized subset group when answering T2 and T3 and reveal that the CF group got the highest average confidence compared to the other two groups for both tasks.
This finding is consistent with the above analysis of these tasks where we found counterfactual visualizations would lead to better performance in T2 and T3.
It also indicates users' average confidence for T2 is higher than T3 which is consistent with the proposed causality comprehension model, implying that T3 requires a higher level of comprehension than T2 (\nameref{sec-model}).
However, our users also provided additional feedback about how counterfactuals may sometimes reduce their original confidence, which is consistent with~\cite{kaul2021improving}. See \nameref{sec-uncertainty} for details.

\autoref{fig:confidence-vis} (c) and (d) indicate that users' confidence for both T2 and T3 is lower for scatterplots compared to bar and line charts.
This finding is consistent with previous work investigating causality illusions, which found that users provided weaker causality ratings for scatterplots than bar charts~\cite{xiong2019illusion}.
However, this finding shows that high user confidence did not necessarily indicate improved performance and this phenomenon needs to be studied in more detail.

\begin{figure}[htbp]
\centering
\includegraphics[width=1\columnwidth]{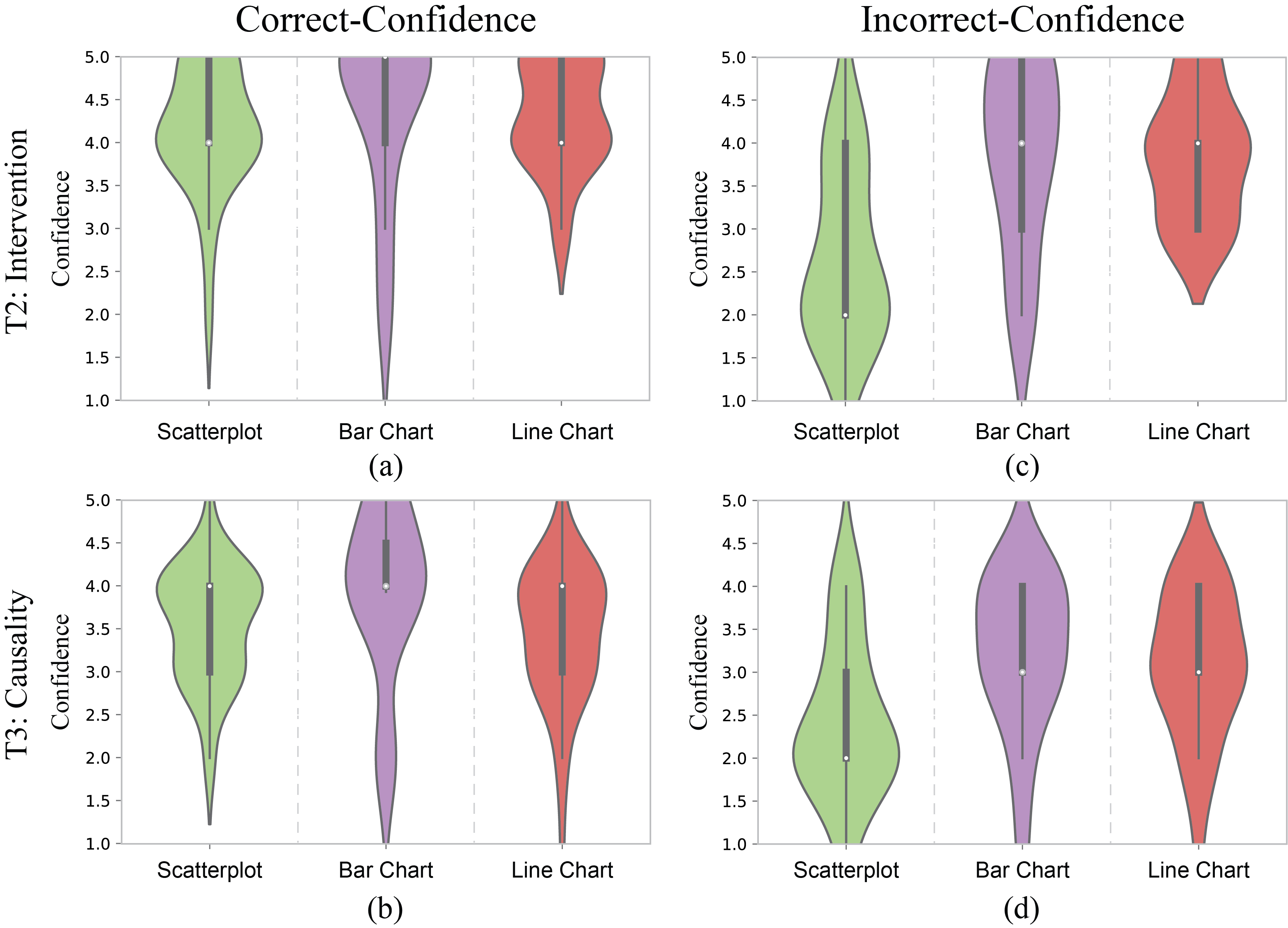}
\caption{The violin plots show the correct-confidence per chart type for T2 (a) and T3 (b) and the incorrect-confidence per chart type for T2 (c) and T3 (d).
}
\label{fig:confidence-correct}
\end{figure}

To better explore the inverse relationship between users' confidence and performance, we further computed and evaluated correct-confidence and incorrect-confidence to measure how users' confidence aligns with their correct and incorrect responses.
The correct-confidence is users' confidence in correct responses while incorrect-confidence refers to confidence in incorrect responses, as shown in \autoref{fig:confidence-correct}.
The results indicate that the differences in correct-confidence in the three chart types do not vary much (see \autoref{fig:confidence-correct} (a) and (b)).
However, the incorrect-confidence of scatterplots is noticeably lower than the other two chart types (see \autoref{fig:confidence-correct} (c) and (d)).
This finding reveals that the relatively low confidence of scatterplots might be explained by correctly having low confidence in incorrect causal inferences.
Thus scatterplots may be positively correlated to the strength of causal evidence behind data, i.e., users would have higher confidence for more significant causal evidence and lower confidence for less significant causal evidence.
These impacts may also be related to the aggregation level of visual design~\cite{xiong2019illusion} which was not considered in our study.
Future work should explore these differences in more detail.

\subsection{User Feedback}
\label{sec-uncertainty}

In addition, we reported a crucial insight into decision-making uncertainty from participants' feedback.
By summarizing participant feedback regarding chart-reading strategies, most users reported that counterfactuals were helpful in finding implicit causal relationships and reasoning about hypothetical scenarios.
However, we also found potential limitations to counterfactuals.

\begin{figure}[htbp]
\centering
\includegraphics[width=0.7\columnwidth]{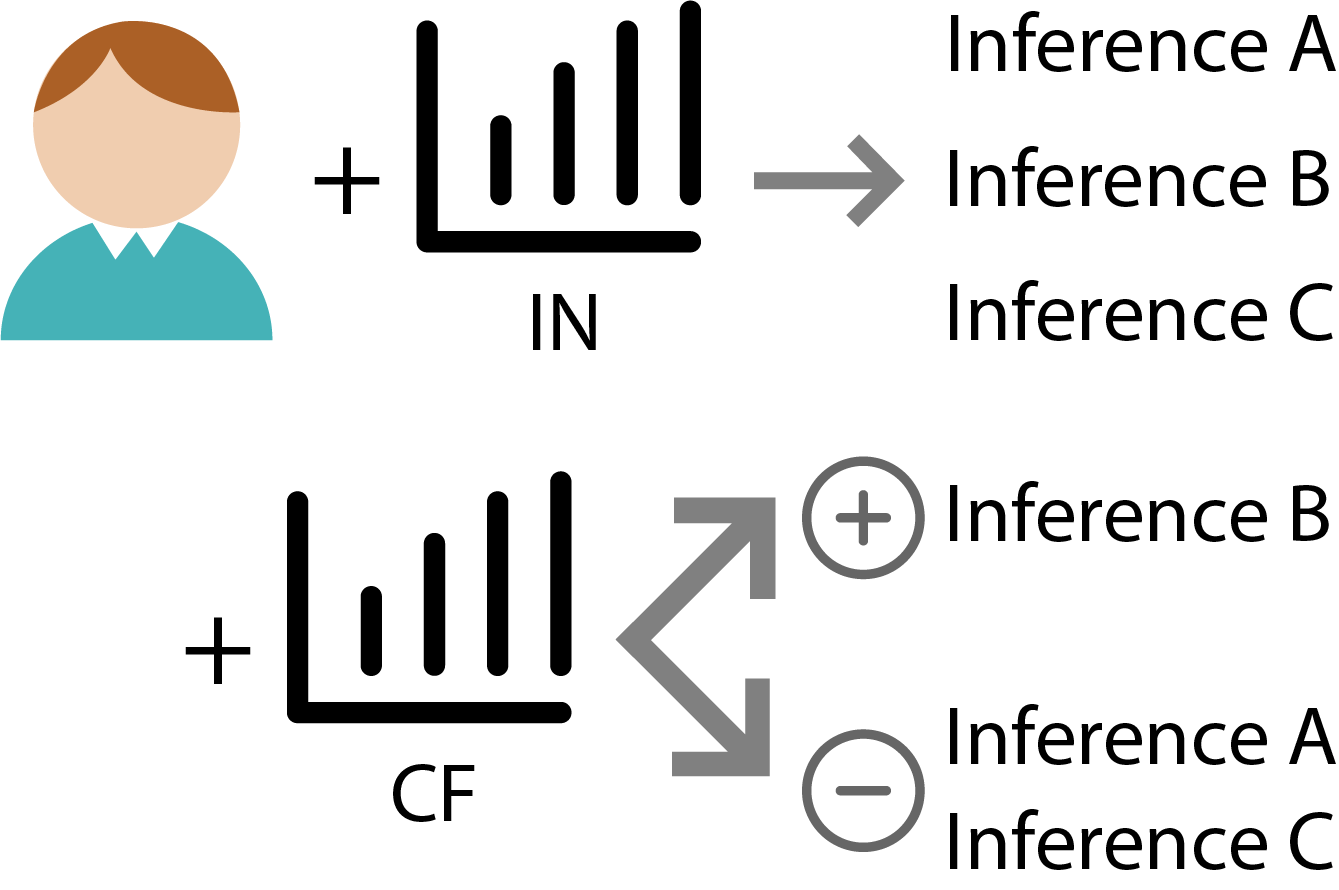}
\caption{Proposed uncertainty-aware decision-making flow based on participants' reports.
The top arrow shows that users may make three inferences A, B, and C when looking at the IN chart, but may not immediately determine which one is the most likely.
The bottom arrow indicates that after looking at the CF charts, it could be possible that the strength of users' inference B is increased while inferences A and C are decreased.
}
\label{fig:feedback}
\end{figure}

\autoref{fig:feedback} shows a conceptual model illustrating how a number of participants described their decision-making strategy and process.
During the study, participants might generate multiple original inferences when looking at the IN chart, where they cannot verify which is correct, but may make an assumption or have a preference among these inferences.
In most cases, when looking at the CF chart, they are then able to confirm or reject the original assumptions.
However, users felt that sometimes the CF chart might ``muddy the waters,'' adding additional information that could be difficult to reconcile with their previous assumptions, and potentially leading to confusion and lack of confidence.
Additionally, users also mentioned that the current study lacks explorations of the whole dataset, due to visualizing sets of static charts, which made their decision-making more difficult.

This situation is similar to the impact of users' decision-making uncertainty found in previous research~\cite{lipshitz1997coping, manski2020lure, kale2019decision}, while an ideal causal inference process should be able to convey the most correct possible decision and reduce users' uncertainty.
The proposed conceptual model could also be helpful to explain how counterfactuals can counter Simpson's paradox in one-dimensional datasets~\cite{wang2023countering}, in which case users may have only one inference initially, and visualizing counterfactuals can counter the inference that a paradox exists.
However, since the evidence for causality is uncertain, increasing uncertainty may also be desirable.
Thus, further analysis is necessary to understand the cognitive processes involved in decision-making in the context of uncertainty and causal inference.

In summary, our study results offer evidence largely confirming our proposed hypotheses, and provide preliminary findings of the impact of visualizing counterfactuals, thus providing insights to support the proposed causality comprehension model in \nameref{sec-model}.

\section{Discussion}
\label{sec-discussion}

Our study primarily evaluated the impact of counterfactual visualizations in helping people understand data at different communication levels.
Our results offer a new perspective on findings from prior studies and provide preliminary design guidance and actionable insights for future research.

\subsection{Critical Reflection Within the Context of Prior Studies}

Our study demonstrates that integrating counterfactual information with visualization can significantly improve users' interpretation of complex datasets, enabling them to operate at a higher data communication and causal inference level.
Next, we discuss the connections between our reported results and insights from prior studies.

Our first major finding reported that counterfactuals do no harm with respect to users' ability to recognize relevant features in visualizations.
This is consistent with Kaul et al.'s study ~\cite{kaul2021improving} where they showed that counterfactuals neither decreased the performance of a visual analytic system nor did they negatively impact user experience.
Similarly, Kale et al.~\cite{kale2021causal} did not find obvious differences in performance when conducting causal inferences with different charts.

Further, our results demonstrate that counterfactuals can significantly help people understand associations and analyze causalities within data.
This finding can be connected with some related insights.
Kaul et al.~\cite{kaul2021improving} found that counterfactual visualizations significantly impact users' inferences drawn from charts.
Our results confirm such beneficial impacts.
Xiong et al.~\cite{xiong2019illusion} found that for the same dataset, different visualization choices could result in differences in users' causal inferences, and that some visualizations could trigger stronger causal relations.
Our results confirm that cognitive reasoning affordances vary for both visualized subsets and chart types, and visualizing counterfactuals can improve users' causal reasoning.

Additionally, our findings indicate that visualizing counterfactuals can help with dataset recall.
By treating counterfactual information as additions to the original visualization, this finding could be consistent with the guidelines from Borkin et al.~\cite{borkin2015beyond}, in which they found that additional encodings can improve the effectiveness of visual data communication.

We also reported specific insights about task completion time, chart types, and users' confidence.
Our finding about completion time confirmed the previous assumption that counterfactual visualizations could be more complex for users to understand~\cite{kaul2021improving}.
This finding also fits with traditional graphical perception problems~\cite{tukey1977exploratory, cleveland1984graphical, heer2010crowdsourcing}, where users spend more time understanding as chart complexity increases. 
However, this may not necessarily be a negative if taking a longer time leads to more correct inferences, as indicated by existing work on cognitive load and memorability~\cite{hullman2011benefitting}.

Finally, our results also suggest that there may be performance differences among chart types. Overall, scatterplots may afford improved causal inferences vs. bar or line charts.
This finding fits with previous studies showing that scatterplots may better convey causal relationships in datasets~\cite{bergstrom2018scatter}, and can more effectively communicate correlations~\cite{kay2015beyond}.
Combined with the results of both performance and confidence, our results additionally show that compared to bar and line charts, users' confidence with scatterplots was more closely aligned with the strength of causal evidence (i.e., positively correlated to their causal inference performance).

Some existing studies, however, also show that scatterplots might convey more uncertainty and be sensitive to different visual encodings~\cite{chan2013generalized, sarma2022evaluating}, and uncertainty may lead to poor decision-making performances~\cite{hullman2019authors, manski2020lure}.
Similarly, users' feedback reported in \nameref{sec-uncertainty} provided some possibilities about how CF could impact uncertainty in their decision-making.
Our study was not designed to investigate the impact of uncertainties in decision-making, however, so this relationship needs to be further explored.

\subsection{Design Implications for Counterfactual Visualizations}

Compared to previous work~\cite{kaul2021improving}, our results go farther in providing preliminary insights about how visualizing counterfactuals effectively can help with data interpretation.
Additionally, we extended previous work on visual causal inference~\cite{xiong2019illusion, kale2021causal}, connecting causal inference to multiple data communication levels.
In this way, 
our results indicate preliminary \add{guidelines for how to use} counterfactuals:

\begin{itemize}

    \item \textit{Visualize counterfactual subsets to convey causalities in datasets.}

    Our study indicates that people can infer causal relations using counterfactual subset visualization, achieving better performance than showing the IN chart only, as described in \nameref{sec-results}.
    In real-world applications, however, counterfactuals have mostly been conveyed via natural language~\cite{feder2022causal}.
    We therefore recommend that designers consider showing counterfactual subsets of data simultaneously with the originally designed chart, if their design objectives include helping their audience find important causalities in the data.

    \item \textit{Use less scalable visualization types to help convey causality.}

    As shown in \nameref{sec-exp}, our exploratory analysis on chart types suggests that there may be evidence that the scalability of charts might impact causal inference.
    Combined with prior similar findings about the impact of aggregation levels~\cite{xiong2019illusion} and rankings~\cite{kay2015beyond}, we anticipate that users might be able to find more causalities in less scalable chart types.
    Thus, we would recommend designers consider using charts with less scalability if they're unsure about the visualization choices for datasets containing complex causalities.
    Specifically, our results in \nameref{sec-exp} indicate that scatterplots can be positively correlated to the strength of causal evidence within datasets.

\end{itemize}

\subsection{Study Limitations}

Our study was able to evaluate the impact of visualizing counterfactuals for visual data communication using static charts.
Some limitations of the study are discussed below.

First, the hypotheses and task designs of our study are primarily based on the proposed causality comprehension model.
\add{The study results provide some support for this model, but further research is required to explore how broadly the preliminary model can be applied across a wider range of scenarios.}
\add{For example,} although we have connected important insights from causal inference and human cognition in visual data communication, other possible impact factors also exist, such as how the potential uncertainty within visualizing counterfactuals will impact users' decision-making~\cite{hullman2018pursuit, padilla2021uncertain}.

Second, we focused primarily on analyzing whether counterfactuals can help, but do not address what constitutes a ``good'' counterfactual subset.
Different aggregations and selections of data subsets may significantly impact the performance~\cite{zhang2015iterative, gotz2019visual} and introduce potential bias~\cite{borland2020selection, borland2019selection, gotz2016adaptive, borland2018contextual} in exploratory analysis.
We adopted the basic counterfactual generation method from previous work~\cite{kaul2021improving}, and did not explore any parameter adjustments for creating the counterfactual subset, thus limiting our findings with respect to the impact of visualizing different counterfactual subsets.

Meanwhile, our study employed the IN chart as a typical visualization, along with counterfactual (IN + CF + REM) and control (IN + EX) groups as comprehensive views of the whole dataset. However, these visualizations show different numbers of charts which may also impact users' perception results.
\add{Additionally, we studied only three basic static visualization types: bar, line, and scatterplot charts. We did not evaluate user understanding of more complex visual designs, such as interactive dashboards and graph representations such as Bayesian Belief Networks~\cite{lam1994learning} that aim to capture multidimensional probabilistic causal relations}. Such interfaces and data representations can be challenging to evaluate~\cite{sarikaya2018we, gotz2019visualization}, and can include different chart aggregation levels \add{and multidimensional relations}, which might complicate the analysis of perceived causality~\cite{xiong2019illusion}.

Additionally, our evaluation employs measures like correctness rate and a number of findings that are usually applied in low-level task evaluation.
\add{However, a limitation common to our work and most existing studies is the lack of a ground truth of human perception of causal relations among variables within datasets, which makes it unclear whether users really judge correlation or causation.
Unlike most traditional visualization tasks such as identifying the number of classes, a user's causal inference is subjective, which means even for the same pair of variables, different people may have different inference criteria.
}
\add{In addition,} user performance for causal inference could also be evaluated by higher-level measures.
For example, the designers' objectives would significantly impact users' comprehension, such as information aesthetics and clarity~\cite{quispel2018aesthetics}.
Further, many other aspects such as individual cognitive factors and personality psychology would have a crucial influence on high-level visual understanding~\cite{ziemkiewicz2012understanding}.
In addition, our participant recruitment was biased toward users attending university, who may be more familiar with statistics and visualizations than a more general population.

Moreover, Kale et al.~\cite{kale2021causal} reported users' causal inferences with common visualizations do not perform significantly better than those visualizing textual contingency tables.
However, our work and several existing visual analytics systems~\cite{deng2021compass, xie2020visual, cheng2020dece, gomez2020vice, kaul2021improving, guo2023causalvis} show that visualization of counterfactual and causal relations would benefit users' interpretation and analyses of data.
We anticipate that this finding could be impacted by the types of chosen charts and the representation of causal information.
It might also be related to the scalability of charts, where scatterplots usually are less scalable and can represent lower variance~\cite{eick2002visual, sarikaya2017scatterplots, kay2015beyond}.
However, this difference definitely needs to be further studied to get a more concrete answer.

\subsection{Future Opportunities}

Based on the aforementioned limitations, it is necessary to assess counterfactual visualizations with more evaluation. This includes understanding metrics such as decision-making uncertainty (as shown in \nameref{sec-uncertainty}), a wider variety of, and control over, counterfactual subset selection criteria, and more complex visual encodings and chart designs \add{such as network-based representations}.
We plan to conduct further experiments to understand how different parameters for specifying counterfactual subsets would impact users' interpretations.
We hope to design different counterfactual visualization techniques that can the show same amount of data samples but with different numbers of charts to further explore the impact of the number of charts on users' perceptions in the future.
We also would like to design more reliable evaluation measures, considering designers' objectives, and extend our study to include a broader population.

\add{In addition, to provide clearer justifications for participants' interpretations of causal relations within datasets, we further plan to conduct a large-scale study judging common inference results across diverse populations for causal inference questions.
Results from such studies could potentially provide a ground truth corpus of causal relations of data variables that would not only benefit counterfactual studies, but also work as baselines for more diverse causality-related empirical experiments.} 

Further, according to our current results, it would be reasonable to assume that visualizing counterfactuals can significantly benefit people's ability to conduct exploratory data analysis.
As a consequence, we hope to explore how users will use and interpret counterfactuals and whether they fit our causality comprehension model for exploratory tasks in interactive visualization systems.
Additionally, our study focused on how users can find causalities, but lacked an understanding of how counterfactuals can guard against making false assumptions of causality, although this is hinted at by the incorrect-confidence result for scatterplots in \nameref{sec-exp}.
Future work should explore this area of counterfactual visualization more thoroughly.

Our results suggest that low-level data communication is not obviously impacted by counterfactuals, as all subset visualization groups achieved an overall high accuracy rate in T1. 
However, existing findings indicate that demographics can influence the accuracy of understanding associations, such as in climate change visual analytics~\cite{ballantyne2016images}.
Future work could further investigate the impact of users' demographics in understanding counterfactuals for more complicated tasks with lower overall correctness rates.

In addition, visualization recommendation and insight characterization methods~\cite{gotz2009behavior, gotz2008characterizing, moritz2018formalizing} have already been a fruitful and insightful research topic. However, unlike with NLP models, it remains difficult to apply those methods in real-world applications~\cite{kaur2017review, zeng2021evaluation}.
In the future, we would like to extend counterfactuals into more complex application scenarios, distill empirically supported counterfactual generation methods, and explore the possibilities for applying counterfactuals in visualization recommendations to provide causality-enhanced insights.

\section{Conclusion}
\label{sec-conclusion}

In this paper, we proposed a method to model the comprehension of causalities from visualizations by combining causal inference theory and cognitive processes of the visual data communication framework.
We explored how counterfactuals impacted people's ability to understand data at different levels for static visualizations via a user study.
Our results indicate that people can interpret and infer relations with counterfactuals.
We provide preliminary evidence that visualizing counterfactuals can improve performance in understanding interventions, analyzing causalities, and recalling features of datasets.
Based on the results evaluation, we discussed the connections and reflections between our results and prior findings to explore more insights.
We further derived design implications for using counterfactuals in visualizations.
We believe our findings could benefit a broad range of visual comprehension demands and tasks, and we hope our work will inform further studies to explore further detailed guidance on how to use and interpret counterfactuals.

\begin{acks}
This work is supported in part by Award \#2211845 from the National Science Foundation.
\end{acks}

\bibliographystyle{SageV}

\begin{thebibliography}{10}
\providecommand{\url}[1]{\texttt{#1}}
\providecommand{\urlprefix}{URL }
\expandafter\ifx\csname urlstyle\endcsname\relax
  \providecommand{\doi}[1]{DOI:\discretionary{}{}{}#1}\else
  \providecommand{\doi}{DOI:\discretionary{}{}{}\begingroup \urlstyle{rm}\Url}\fi
\providecommand{\eprint}[2][]{\url{#2}}

\bibitem{gehlenborg2010visualization}
Gehlenborg N, O'Donoghue SI, Baliga NS et~al.
\newblock Visualization of omics data for systems biology.
\newblock \emph{Nature Methods} 2010; 7(Suppl 3): S56--S68.

\bibitem{kong2018frames}
Kong HK, Liu Z and Karahalios K.
\newblock Frames and slants in titles of visualizations on controversial topics.
\newblock In \emph{ACM SIGCHI Conference on Human Factors in Computing Systems}. pp. 1--12.

\bibitem{walny2019data}
Walny J, Frisson C, West M et~al.
\newblock Data changes everything: Challenges and opportunities in data visualization design handoff.
\newblock \emph{IEEE Transactions on Visualization and Computer Graphics} 2019; 26(1): 12--22.

\bibitem{wang2015visual}
Wang J and Mueller K.
\newblock The visual causality analyst: An interactive interface for causal reasoning.
\newblock \emph{IEEE Transactions on Visualization and Computer Graphics} 2015; 22(1): 230--239.

\bibitem{wang2017visual}
Wang J and Mueller K.
\newblock Visual causality analysis made practical.
\newblock In \emph{2017 IEEE Conference on Visual Analytics Science and Technology (VAST)}. IEEE, pp. 151--161.

\bibitem{pearl2009causal}
Pearl J.
\newblock {Causal inference in statistics: An overview}.
\newblock \emph{Statistics Surveys} 2009; 3(none): 96 -- 146.

\bibitem{pearl2009causality}
Pearl J.
\newblock \emph{Causality}.
\newblock Cambridge university press, 2009.

\bibitem{cheng2020dece}
Cheng F, Ming Y and Qu H.
\newblock Dece: Decision explorer with counterfactual explanations for machine learning models.
\newblock \emph{IEEE Transactions on Visualization and Computer Graphics} 2020; 27(2): 1438--1447.

\bibitem{gomez2020vice}
Gomez O, Holter S, Yuan J et~al.
\newblock Vice: Visual counterfactual explanations for machine learning models.
\newblock In \emph{Proceedings of the 25th International Conference on Intelligent User Interfaces}. pp. 531--535.

\bibitem{kaul2021improving}
Kaul S, Borland D, Cao N et~al.
\newblock Improving visualization interpretation using counterfactuals.
\newblock \emph{IEEE Transactions on Visualization and Computer Graphics} 2021; 28(1): 998--1008.

\bibitem{xiong2019illusion}
Xiong C, Shapiro J, Hullman J et~al.
\newblock Illusion of causality in visualized data.
\newblock \emph{IEEE Transactions on Visualization and Computer Graphics} 2019; 26(1): 853--862.

\bibitem{kale2021causal}
Kale A, Wu Y and Hullman J.
\newblock Causal support: modeling causal inferences with visualizations.
\newblock \emph{IEEE Transactions on Visualization and Computer Graphics} 2021; 28(1): 1150--1160.

\bibitem{adar2020communicative}
Adar E and Lee E.
\newblock Communicative visualizations as a learning problem.
\newblock \emph{IEEE Transactions on Visualization and Computer Graphics} 2020; 27(2): 946--956.

\bibitem{ajani2021declutter}
Ajani K, Lee E, Xiong C et~al.
\newblock Declutter and focus: Empirically evaluating design guidelines for effective data communication.
\newblock \emph{IEEE Transactions on Visualization and Computer Graphics} 2021; 28(10): 3351--3364.

\bibitem{lee2022affective}
Lee-Robbins E and Adar E.
\newblock Affective learning objectives for communicative visualizations.
\newblock \emph{IEEE Transactions on Visualization and Computer Graphics} 2022; 29(1): 1--11.

\bibitem{bergstrom2018scatter}
Bergstrom CT and West JD.
\newblock Why scatter plots suggest causality, and what we can do about it.
\newblock \emph{arXiv preprint arXiv:180909328} 2018; .

\bibitem{shneiderman1996eyes}
Shneiderman B.
\newblock The eyes have it: A task by data type taxonomy for information visualizations.
\newblock In \emph{Proceedings 1996 IEEE Symposium on Visual Languages}. IEEE, pp. 336--343.

\bibitem{tukey1962future}
Tukey JW.
\newblock The future of data analysis.
\newblock \emph{The Annals of Mathematical Statistics} 1962; 33(1): 1--67.

\bibitem{tukey1977exploratory}
Tukey JW et~al.
\newblock \emph{Exploratory data analysis}, volume~2.
\newblock Reading, MA, 1977.

\bibitem{van2005value}
Van~Wijk JJ.
\newblock The value of visualization.
\newblock In \emph{IEEE Visualization}. pp. 79--86.

\bibitem{munzner2014visualization}
Munzner T.
\newblock \emph{Visualization analysis and design}.
\newblock CRC press, 2014.

\bibitem{cleveland1984graphical}
Cleveland WS and McGill R.
\newblock Graphical perception: Theory, experimentation, and application to the development of graphical methods.
\newblock \emph{Journal of the American Statistical Association} 1984; 79(387): 531--554.

\bibitem{heer2008graphical}
Heer J, Mackinlay J, Stolte C et~al.
\newblock Graphical histories for visualization: Supporting analysis, communication, and evaluation.
\newblock \emph{IEEE Transactions on Visualization and Computer Graphics} 2008; 14(6): 1189--1196.

\bibitem{szafir2018modeling}
Szafir DA.
\newblock Modeling color difference for visualization design.
\newblock \emph{IEEE Transactions on Visualization and Computer Graphics} 2018; 24(1): 392--401.
\newblock \doi{10.1109/TVCG.2017.2744359}.

\bibitem{tseng2023measuring}
Tseng C, Quadri GJ, Wang Z et~al.
\newblock Measuring categorical perception in color-coded scatterplots.
\newblock In \emph{proceedings of the 2023 CHI Conference on Human Factors in Computing Systems}. pp. 1--14.

\bibitem{sedlmair2012design}
Sedlmair M, Meyer M and Munzner T.
\newblock Design study methodology: Reflections from the trenches and the stacks.
\newblock \emph{IEEE Transactions on Visualization and Computer Graphics} 2012; 18(12): 2431--2440.

\bibitem{sedlmair2015data}
Sedlmair M and Aupetit M.
\newblock Data-driven evaluation of visual quality measures.
\newblock \emph{Computer Graphics Forum} 2015; 34.
\newblock \doi{10.1111/cgf.12632}.

\bibitem{wilkinson2005graph}
Wilkinson L, Anand A and Grossman R.
\newblock Graph-theoretic scagnostics.
\newblock In \emph{IEEE Symposium on Information Visualization (InfoVis)}. IEEE, pp. 157--164.
\newblock \doi{10.1109/INFVIS.2005.1532142}.

\bibitem{wang2019improving}
Wang Y, Wang Z, Liu T et~al.
\newblock Improving the robustness of scagnostics.
\newblock \emph{IEEE Transactions on Visualization and Computer Graphics} 2019; 26(1): 759--769.

\bibitem{bae2022cultivating}
Bae SS, Vanukuru R, Yang R et~al.
\newblock Cultivating visualization literacy for children through curiosity and play.
\newblock \emph{IEEE Transactions on Visualization and Computer Graphics} 2022; 29(1): 257--267.

\bibitem{bloom2020taxonomy}
Bloom BS and Krathwohl DR.
\newblock \emph{Taxonomy of educational objectives: The classification of educational goals. Book 1, Cognitive domain}.
\newblock longman, 2020.

\bibitem{wu2021polyjuice}
Wu T, Ribeiro MT, Heer J et~al.
\newblock Polyjuice: Generating counterfactuals for explaining, evaluating, and improving models.
\newblock In \emph{Proceedings of the 59th Annual Meeting of the Association for Computational Linguistics}. pp. 6707--6723.

\bibitem{qin2019counterfactual}
Qin L, Bosselut A, Holtzman A et~al.
\newblock Counterfactual story reasoning and generation.
\newblock In \emph{Proceedings of the 2019 Conference on Empirical Methods in Natural Language Processing}. pp. 5043--5053.

\bibitem{borland2024using}
Borland D, Wang AZ and Gotz D.
\newblock Using Counterfactuals to Improve Causal Inferences from Visualizations.
\newblock \emph{IEEE Computer Graphics and Applications} 2024; 44(1).
\newblock \doi{10.1109/MCG.2023.3338788}.

\bibitem{wexler2019if}
Wexler J, Pushkarna M, Bolukbasi T et~al.
\newblock The what-if tool: Interactive probing of machine learning models.
\newblock \emph{IEEE Transactions on Visualization and Computer Graphics} 2019; 26(1): 56--65.

\bibitem{ciorna2023interact}
Ciorna V, Melançon G, Petry F et~al.
\newblock Interact: A visual what-if analysis tool for virtual product design.
\newblock \emph{Information Visualization} 2023.
\newblock \doi{10.1177/14738716231216030}.

\bibitem{oghbaie2016understanding}
Oghbaie M, Pennock MJ and Rouse WB.
\newblock Understanding the efficacy of interactive visualization for decision making for complex systems.
\newblock In \emph{2016 Annual IEEE Systems Conference (SysCon)}. IEEE, pp. 1--6.

\bibitem{griffiths2005structure}
Griffiths TL and Tenenbaum JB.
\newblock Structure and strength in causal induction.
\newblock \emph{Cognitive Psychology} 2005; 51(4): 334--384.

\bibitem{deng2021compass}
Deng Z, Weng D, Xie X et~al.
\newblock Compass: Towards better causal analysis of urban time series.
\newblock \emph{IEEE Transactions on Visualization and Computer Graphics} 2021; 28(1): 1051--1061.

\bibitem{xie2020visual}
Xie X, Du F and Wu Y.
\newblock A visual analytics approach for exploratory causal analysis: Exploration, validation, and applications.
\newblock \emph{IEEE Transactions on Visualization and Computer Graphics} 2020; 27(2): 1448--1458.

\bibitem{guo2023causalvis}
Guo G, Karavani E, Endert A et~al.
\newblock Causalvis: Visualizations for causal inference.
\newblock In \emph{Proceedings of the 2023 CHI Conference on Human Factors in Computing Systems}. pp. 1--20.

\bibitem{lam1994learning}
Lam W and Fahiem B.
\newblock Learning Bayesian belief networks: An approach based on the MDL principle.
\newblock \emph{Computational Intelligence} 1994; 10(3): 269--293.

\bibitem{plaisant2004challenge}
Plaisant C.
\newblock The challenge of information visualization evaluation.
\newblock In \emph{Proceedings of the 2004 working conference on Advanced visual interfaces}. pp. 109--116.

\bibitem{lam2011empirical}
Lam H, Bertini E, Isenberg P et~al.
\newblock Empirical studies in information visualization: Seven scenarios.
\newblock \emph{IEEE Transactions on Visualization and Computer Graphics} 2011; 18(9): 1520--1536.

\bibitem{borkin2015beyond}
Borkin MA, Bylinskii Z, Kim NW et~al.
\newblock Beyond memorability: Visualization recognition and recall.
\newblock \emph{IEEE Transactions on Visualization and Computer Graphics} 2015; 22(1): 519--528.

\bibitem{kong2019trust}
Kong HK, Liu Z and Karahalios K.
\newblock Trust and recall of information across varying degrees of title-visualization misalignment.
\newblock In \emph{Proceedings of the 2019 CHI Conference on Human Factors in Computing Systems}. pp. 1--13.

\bibitem{faul2007g}
Faul F, Erdfelder E, Lang AG et~al.
\newblock G* power 3: A flexible statistical power analysis program for the social, behavioral, and biomedical sciences.
\newblock \emph{Behavior Research Methods} 2007; 39(2): 175--191.

\bibitem{amar2005low}
Amar R, Eagan J and Stasko J.
\newblock Low-level components of analytic activity in information visualization.
\newblock In \emph{IEEE Symposium on Information Visualization (InfoVis)}. pp. 111--117.
\newblock \doi{10.1109/INFVIS.2005.1532136}.

\bibitem{regier2009language}
Regier T and Kay P.
\newblock Language, thought, and color: Whorf was half right.
\newblock \emph{Trends in Cognitive Sciences} 2009; 13(10): 439--446.

\bibitem{quadri2021survey}
Quadri GJ and Rosen P.
\newblock A survey of perception-based visualization studies by task.
\newblock \emph{IEEE Transactions on Visualization and Computer Graphics} 2021; .

\bibitem{gleicher2011visual}
Gleicher M, Albers D, Walker R et~al.
\newblock Visual comparison for information visualization.
\newblock \emph{Information Visualization} 2011; 10(4): 289--309.

\bibitem{zhang2015glyph}
Zhang C, Schultz T, Lawonn K et~al.
\newblock Glyph-based comparative visualization for diffusion tensor fields.
\newblock \emph{IEEE Transactions on Visualization and Computer Graphics} 2015; 22(1): 797--806.

\bibitem{gleicher2017considerations}
Gleicher M.
\newblock Considerations for visualizing comparison.
\newblock \emph{IEEE Transactions on Visualization and Computer Graphics} 2017; 24(1): 413--423.

\bibitem{tang2017extracting}
Tang B, Han S, Yiu ML et~al.
\newblock Extracting top-k insights from multi-dimensional data.
\newblock In \emph{Proceedings of the 2017 ACM International Conference on Management of Data}. pp. 1509--1524.

\bibitem{ding2019quickinsights}
Ding R, Han S, Xu Y et~al.
\newblock Quickinsights: Quick and automatic discovery of insights from multi-dimensional data.
\newblock In \emph{Proceedings of the 2019 International Conference on Management of Data}. pp. 317--332.

\bibitem{ma2021metainsight}
Ma P, Ding R, Han S et~al.
\newblock Metainsight: Automatic discovery of structured knowledge for exploratory data analysis.
\newblock In \emph{Proceedings of the 2021 International Conference on Management of Data}. pp. 1262--1274.

\bibitem{yeh2009comparisons}
Yeh IC and Lien Ch.
\newblock The comparisons of data mining techniques for the predictive accuracy of probability of default of credit card clients.
\newblock \emph{Expert Systems with Applications} 2009; 36(2): 2473--2480.

\bibitem{kohavi1996scaling}
Kohavi R et~al.
\newblock Scaling up the accuracy of naive-bayes classifiers: A decision-tree hybrid.
\newblock In \emph{Proceedings of the 1996 International Conference on Knowledge Discovery and Data Mining}. pp. 202--207.

\bibitem{lin2014microsoft}
Lin TY, Maire M, Belongie S et~al.
\newblock Microsoft coco: Common objects in context.
\newblock In \emph{Computer Vision--ECCV 2014: 13th European Conference, Zurich, Switzerland, September 6-12, 2014, Proceedings, Part V 13}. Springer, pp. 740--755.

\bibitem{charmaz2006constructing}
Charmaz K.
\newblock \emph{Constructing grounded theory: A practical guide through qualitative analysis}.
\newblock sage, 2006.

\bibitem{dragicevic2016fair}
Dragicevic P.
\newblock Fair statistical communication in hci.
\newblock \emph{Modern Statistical Methods for HCI} 2016; : 291--330.

\bibitem{owidco2andgreenhousegasemissions}
Ritchie H, Roser M and Rosado P.
\newblock Co2 and greenhouse gas emissions.
\newblock \emph{Our World in Data} 2020.

\bibitem{world2021ghe}
Organization WH et~al.
\newblock Ghe: Life expectancy and healthy life expectancy.
\newblock \emph{The Global Health Observatory} 2021.

\bibitem{lipshitz1997coping}
Lipshitz R and Strauss O.
\newblock Coping with uncertainty: A naturalistic decision-making analysis.
\newblock \emph{Organizational Behavior and Human Decision Processes} 1997; 69(2): 149--163.

\bibitem{manski2020lure}
Manski CF.
\newblock The lure of incredible certitude.
\newblock \emph{Economics \& Philosophy} 2020; 36(2): 216--245.

\bibitem{kale2019decision}
Kale A, Kay M and Hullman J.
\newblock Decision-making under uncertainty in research synthesis: Designing for the garden of forking paths.
\newblock In \emph{Proceedings of the 2019 CHI Conference on Human Factors in Computing Systems}. pp. 1--14.

\bibitem{wang2023countering}
Wang AZ, Borland D and Gotz D.
\newblock Countering simpson’s paradox with counterfactuals.
\newblock In \emph{Poster Proceedings of IEEE VIS}. pp. 1--2.

\bibitem{heer2010crowdsourcing}
Heer J and Bostock M.
\newblock Crowdsourcing graphical perception: using mechanical turk to assess visualization design.
\newblock In \emph{Proceedings of the 2010 CHI Conference on Human Factors in Computing Systems}. pp. 203--212.

\bibitem{hullman2011benefitting}
Hullman J, Adar E and Shah P.
\newblock Benefitting infovis with visual difficulties.
\newblock \emph{IEEE Transactions on Visualization and Computer Graphics} 2011; 17(12): 2213--2222.

\bibitem{kay2015beyond}
Kay M and Heer J.
\newblock Beyond weber's law: A second look at ranking visualizations of correlation.
\newblock \emph{IEEE Transactions on Visualization and Computer Graphics} 2015; 22(1): 469--478.

\bibitem{chan2013generalized}
Chan YH, Correa CD and Ma KL.
\newblock The generalized sensitivity scatterplot.
\newblock \emph{IEEE Transactions on Visualization and Computer Graphics} 2013; 19(10): 1768--1781.

\bibitem{sarma2022evaluating}
Sarma A, Guo S, Hoffswell J et~al.
\newblock Evaluating the use of uncertainty visualisations for imputations of data missing at random in scatterplots.
\newblock \emph{IEEE Transactions on Visualization and Computer Graphics} 2022; 29(1): 602--612.

\bibitem{hullman2019authors}
Hullman J.
\newblock Why authors don't visualize uncertainty.
\newblock \emph{IEEE Transactions on Visualization and Computer Graphics} 2019; 26(1): 130--139.

\bibitem{feder2022causal}
Feder A, Keith KA, Manzoor E et~al.
\newblock Causal inference in natural language processing: Estimation, prediction, interpretation and beyond.
\newblock \emph{Transactions of the Association for Computational Linguistics} 2022; 10: 1138--1158.

\bibitem{hullman2018pursuit}
Hullman J, Qiao X, Correll M et~al.
\newblock In pursuit of error: A survey of uncertainty visualization evaluation.
\newblock \emph{IEEE Transactions on Visualization and Computer Graphics} 2018; 25(1): 903--913.

\bibitem{padilla2021uncertain}
Padilla LM, Powell M, Kay M et~al.
\newblock Uncertain about uncertainty: How qualitative expressions of forecaster confidence impact decision-making with uncertainty visualizations.
\newblock \emph{Frontiers in Psychology} 2021; 11: 579267.

\bibitem{zhang2015iterative}
Zhang Z, Gotz D and Perer A.
\newblock Iterative cohort analysis and exploration.
\newblock \emph{Information Visualization} 2015; 14(4): 289--307.

\bibitem{gotz2019visual}
Gotz D, Zhang J, Wang W et~al.
\newblock Visual analysis of high-dimensional event sequence data via dynamic hierarchical aggregation.
\newblock \emph{IEEE Transactions on Visualization and Computer Graphics} 2019; 26(1): 440--450.

\bibitem{borland2020selection}
Borland D, Zhang J, Kaul S et~al.
\newblock Selection-bias-corrected visualization via dynamic reweighting.
\newblock \emph{IEEE Transactions on Visualization and Computer Graphics} 2020; 27(2): 1481--1491.

\bibitem{borland2019selection}
Borland D, Wang W, Zhang J et~al.
\newblock Selection bias tracking and detailed subset comparison for high-dimensional data.
\newblock \emph{IEEE Transactions on Visualization and Computer Graphics} 2019; 26(1): 429--439.

\bibitem{gotz2016adaptive}
Gotz D, Sun S and Cao N.
\newblock Adaptive contextualization: Combating bias during high-dimensional visualization and data selection.
\newblock In \emph{Proceedings of the 21st International Conference on Intelligent User Interfaces}. pp. 85--95.

\bibitem{borland2018contextual}
Borland D, Wang W and Gotz D.
\newblock Contextual visualization.
\newblock \emph{IEEE Computer Graphics and Applications} 2018; 38(6): 17--23.

\bibitem{sarikaya2018we}
Sarikaya A, Correll M, Bartram L et~al.
\newblock What do we talk about when we talk about dashboards?
\newblock \emph{IEEE Transactions on Visualization and Computer Graphics} 2018; 25(1): 682--692.

\bibitem{gotz2019visualization}
Gotz D, Wang W, Chen AT et~al.
\newblock Visualization model validation via inline replication.
\newblock \emph{Information Visualization} 2019; 18(4): 405--425.

\bibitem{quispel2018aesthetics}
Quispel A, Maes A and Schilperoord J.
\newblock Aesthetics and clarity in information visualization: The designer’s perspective.
\newblock In \emph{Arts}, volume~7. MDPI, p.~72.

\bibitem{ziemkiewicz2012understanding}
Ziemkiewicz C, Ottley A, Crouser RJ et~al.
\newblock Understanding visualization by understanding individual users.
\newblock \emph{IEEE Computer Graphics and Applications} 2012; 32(6): 88--94.

\bibitem{eick2002visual}
Eick SG and Karr AF.
\newblock Visual scalability.
\newblock \emph{Journal of Computational and Graphical Statistics} 2002; 11(1): 22--43.

\bibitem{sarikaya2017scatterplots}
Sarikaya A and Gleicher M.
\newblock Scatterplots: Tasks, data, and designs.
\newblock \emph{IEEE Transactions on Visualization and Computer Graphics} 2017; 24(1): 402--412.

\bibitem{ballantyne2016images}
Ballantyne AG, Wibeck V and Neset TS.
\newblock Images of climate change--a pilot study of young people’s perceptions of ict-based climate visualization.
\newblock \emph{Climatic change} 2016; 134: 73--85.

\bibitem{gotz2009behavior}
Gotz D and Wen Z.
\newblock Behavior-driven visualization recommendation.
\newblock In \emph{Proceedings of the 14th international conference on Intelligent user interfaces}. pp. 315--324.

\bibitem{gotz2008characterizing}
Gotz D and Zhou MX.
\newblock Characterizing users’ visual analytic activity for insight provenance.
\newblock In \emph{2008 IEEE Symposium on Visual Analytics Science and Technology}. IEEE, pp. 123--130.

\bibitem{moritz2018formalizing}
Moritz D, Wang C, Nelson GL et~al.
\newblock Formalizing visualization design knowledge as constraints: Actionable and extensible models in draco.
\newblock \emph{IEEE Transactions on Visualization and Computer Graphics} 2018; 25(1): 438--448.

\bibitem{kaur2017review}
Kaur P and Owonibi M.
\newblock A review on visualization recommendation strategies.
\newblock In \emph{International Conference on Information Visualization Theory and Applications}, volume~4. SCITEPRESS, pp. 266--273.

\bibitem{zeng2021evaluation}
Zeng Z, Moh P, Du F et~al.
\newblock An evaluation-focused framework for visualization recommendation algorithms.
\newblock \emph{IEEE Transactions on Visualization and Computer Graphics} 2021; 28(1): 346--356.

\end{thebibliography}

\end{document}